\documentclass[journal]{IEEEtran}

\usepackage[colorlinks=true, linkcolor=blue]{hyperref} 
\usepackage{amsmath,amsfonts}
\usepackage{algorithmic}
\usepackage{algorithm}
\usepackage{array}
\usepackage[caption=false,font=normalsize,labelfont=sf,textfont=sf]{subfig}
\usepackage{textcomp}
\usepackage{stfloats}
\usepackage{url}
\usepackage{verbatim}
\usepackage{graphicx}

\usepackage{xcolor}
\usepackage{makecell}
\usepackage{framed}
\usepackage{booktabs}
\usepackage{longtable}

\usepackage[most]{tcolorbox}

\usepackage{ragged2e}
\usepackage{changepage}
\usepackage{multirow}
\usepackage[caption=false]{subfig}

\usepackage{xcolor}
\usepackage{xspace}

\usepackage{float}
\usepackage{newfloat}
\newfloat{listing}{tbp}{lol}
\floatname{listing}{Code}

\definecolor{diffadd}{HTML}{E6FFED} 
\definecolor{diffdel}{HTML}{FFECEC} 
\definecolor{diffnone}{HTML}{FFFFFF}

\newcommand{\MyBox}[1]{%
    \par
    \addvspace{3pt}
    \noindent\fbox{
        \parbox[t]{.95\columnwidth}{
            \fontsize{10.0}{9}\selectfont
            \justifying #1
        }
    }
}

\newcommand\footnoteref[1]
{\protected@xdef\@thefnmark{\ref{#1}}\@footnotemark}
\makeatother

\def\BibTeX{{\rm B\kern-.05em{\sc i\kern-.025em b}\kern-.08em
    T\kern-.1667em\lower.7ex\hbox{E}\kern-.125emX}}

\newcommand{\responsetoreviewer}[3][black]{

\expandafter\gdef\csname #2\endcsname{#3}%
\label{rev:#2}%
{\color{blue}{#3}}\color{#1}\xspace}
\makeatother

\begin{document}

\title{On the Influence of Refactoring Types on Merge Effort}

\author{
André Oliveira, João Victor Monteiro, Vânia Neves, Alexandre Plastino, Bianca Trinkenreich, Alessandro Garcia~\IEEEmembership{Member, IEEE}, and Leonardo Murta

\thanks{André Oliveira, Vânia Neves, Alexandre Plastino, and Leonardo Murta are with the Computing Institute, Fluminense Federal University, Niterói, RJ, Brazil. E-mail: \{andrelucio, vania, plastino, leomurta\}@ic.uff.br.
Bianca Trinkenreich is with the Colorado State University, Fort Collins, United States of America. E-mail: bianca.trinkenreich@colostate.edu.
Alessandro Garcia is with the Pontifical Catholic University of Rio de Janeiro (PUC-Rio), Rio de Janeiro, Brazil. E-mail: afgarcia@inf.puc-rio.br.
João Victor Monteiro is with the Veiga de Almeida University. E-mail: jjoaovictormonteiro@gmail.com.}

}

\markboth{Journal of \LaTeX\ Class Files,~Vol.~XX, No.~X, Month~202X}%
{Shell \MakeLowercase{\textit{et al.}}: A Sample Article Using IEEEtran.cls for IEEE Journals}

\maketitle

\begin{abstract}
Modern software development involves parallel work and concurrent changes, requiring code merging. Prior studies report that 10\% to 20\% of merge attempts result in conflicts, often requiring manual intervention. The literature explores factors that generate conflicts, including refactorings, but does not analyze how individual refactoring types influence the manual effort required to resolve them. We analyzed 64 open-source Java projects and applied association rule mining to measure the strength of associations between specific refactoring types and merge effort. Our results show that refactoring types relate to merge effort with varying strength. In particular, \textit{Rename Attribute}, \textit{Move Class}, \textit{Extract Variable}, \textit{Change Return Type}, and \textit{Split Parameter} exhibit some of the strongest associations, especially when a higher number of such refactorings is present in the merge branches. We also find that both the number of refactorings and their diversity independently increase merge effort, both in terms of occurrence and intensity. Additionally, the co-occurrence of refactorings across parallel branches is associated with higher merge effort, particularly when combining structural transformations with changes to method signatures and data-structure representations, whereas more localized changes are less frequent in the most impactful combinations.
\end{abstract}


\section{Introduction}
\label{sec:introduction}

During software development, teams make concurrent modifications. They must merge these changes, a challenging process~\cite{ghiotto2020}. Studies report 10\%--20\% of merges failing~\cite{brun2011, kasi2013}, with some projects approaching 50\%~\cite{brun2011, zimmermann2007}. Merging parallel changes requires considerable effort because teams must resolve conflicts and therefore decrease software productivity. Researchers have proposed multiple techniques to resolve merge conflicts~\cite{mens2002, Apel2011}, including automated~\cite{shen2005, apel2012, lebenich2015} and semi-automated methods~\cite{mens2002, apiwattanapong2007, Apel2011}. However, developers often resolve conflicts manually when tools fail~\cite{ghiotto2020}. In any case, given the limited accuracy of existing tools, developers still need to verify merge resolutions manually.

Developers modify the source code for several well-established reasons. They fix bugs (corrective maintenance), adapt the system to new environments (adaptive maintenance), improve performance, testability, security, and usability (perfective maintenance), and prepare the system for future changes (preventive maintenance)~\cite{albuquerque2023, charoenwet2024}. Permeating these types of code changes, we have refactorings that help manage software evolution while preserving design quality. They restructure code without changing behavior, improving modularity and maintainability~\cite{tsantalis2016, fowler2018, pantiuchina2020}. Depending on volume and type, refactorings can affect many code regions. Developers must weigh short-term impacts, as unmanaged refactorings may overlap with parallel changes and trigger merge conflicts.

To illustrate how refactorings contribute to merge effort, we present a real example from GitHub that shows a small excerpt of the \textit{diff}. Code~\ref{cod:MergeEffortExample2} shows a scenario from merge commit \texttt{77e002a} in the \textit{byte-buddy} repository\footnote{\url{https://github.com/raphw/byte-buddy}}, obtained by the \textit{git diff  77e002a  77e002a\^{}1 77e002a\^{}2} command. Here, \texttt{77e002a\^{}1} and \texttt{77e002a\^{}2} denote the first and second parents of the merge commit, respectively. The \textit{diff} uses a unified format where ``-'' and ``+'' denote removals and additions in the branches, respectively, while ``\texttt{--}'' and ``\texttt{++}'' indicate edits introduced during merge resolution. In this case, branch~1 applied a \textit{Change Variable Type} refactoring by introducing \textit{MethodList$<$?$>$}, whereas branch~2 performed a \textit{Rename Attribute} refactoring, shortening \mbox{\textit{targetMethodCandidates}} to \textit{targetCandidates}. The merge combines both changes, yielding \textit{MethodList$<$?$>$ targetCandidates}. The presence of a ``\texttt{++}'' line corresponds to one unit of merge effort (\textit{effort} = 1). This example highlights that even simple and seemingly independent refactorings, when applied concurrently, may require manual intervention during merging, motivating the need to better understand how different refactoring types contribute to merge effort.

\begin{listing}[htbp]
\caption{Example of merge effort identified in commit 77e002a of the \textit{byte-buddy} repository.}
\label{cod:MergeEffortExample2}

\begin{lstlisting}[
    basicstyle=\ttfamily\fontsize{6}{7}\selectfont,
    escapeinside=@@,
    columns=fullflexible,
    keepspaces=true
]
@\colorbox{diffdel}{\makebox[\linewidth][l]{\ttfamily\fontsize{6}{7}\selectfont\strut -\ \ \ private final MethodList<?> targetMethodCandidates;}}@
@\colorbox{diffdel}{\makebox[\linewidth][l]{\ttfamily\fontsize{6}{7}\selectfont\strut \ - private final MethodList targetCandidates;}}@
@\colorbox{diffadd}{\makebox[\linewidth][l]{\ttfamily\fontsize{6}{7}\selectfont\strut ++ private final MethodList<?> targetCandidates;}}@
\end{lstlisting}

\end{listing}

Prior studies have analyzed the relationship between refactorings and code merging~\cite{dig2007, lebenich2017, mahmoudi2018, laszlo2007, mahmoudi2019}. Early work proposed tools to detect common refactorings, such as renaming and moving code, before merges to support developers’ decisions~\cite{dig2007, laszlo2007, lebenich2017}. Mahmoudi and Nadi~\cite{mahmoudi2018} investigated frequent refactoring types and the potential for automated merging, while Mahmoudi et~al.~\cite{mahmoudi2019} found that 22\% of merge conflicts involve refactorings and that 11\% of conflicting regions contain at least one refactoring, which makes such conflicts more complex to resolve.  However, these studies did not examine which refactoring types are more likely to increase the frequency and intensity of manual effort required to resolve merge conflicts, nor whether some refactoring types are (in)compatible to occur in parallel, across branches. Moreover, they considered a limited subset of refactorings compared to the catalog described by Fowler~\cite{fowler2018}.

This paper investigates the influence of individual refactoring types on the manual effort required to resolve merge conflicts, referred to as \textit{merge effort}. We analyze 64 open-source Java projects hosted on GitHub, covering 33 distinct refactoring types, and apply association rule mining~\cite{han2022} to extract co-occurrence patterns between these types and the computed merge effort. We adopt code churn as a surrogate for effort, following Sjøberg et al.~\cite{sjoberg2012}, who reported a moderate to strong correlation between churn and maintenance effort.

This study extends our previous work published at ICSE 2023~\cite{oliveira2023}. In that study, we investigated how the overall number of refactorings and their distribution across merge branches influence merge effort. The results showed that merges involving refactorings occurred with a 24\% higher frequency of merge effort, and that concurrent refactorings in both branches were substantially more frequently associated with high-intensity effort. However, that study analyzed refactorings only at an aggregated level, without distinguishing the specific types of refactoring operations involved. In this paper, we extend that analysis by investigating the roles of individual refactoring types and their interactions, and we use RefactoringMiner v3.1, which offers much higher accuracy for refactoring detection than the version used in our prior study. We consider a wide range of 33 refactoring types and analyze how their occurrence, quantity, diversity, and combinations across branches are associated with merge effort. This type-level analysis provides a more fine-grained understanding of how different refactoring operations influence merge effort during code integration.

 To investigate the influence of refactoring types on merge effort, we address the following research questions, detailed in Section~\ref{sec:results}:

\textbf{RQ1}: Which refactoring types most influence the occurrence of merge effort?

\textit{Understanding how different refactoring types influence merge effort is essential for improving collaborative development reliability. This RQ examines whether some refactoring types are more strongly associated with merge effort than others.}

\textbf{RQ2}: Does the amount of different refactoring types influence the merge effort?

\textit{The analysis in RQ1 considers just the occurrence or not of each refactoring type. However, for the cases where a given refactoring type occurs, the amount of refactorings of such type may also reveal new risk factors in code integration.}

\textbf{RQ3}: How do combinations of refactoring types, applied concurrently in different branches, influence the merge effort?

\textit{Refactorings implemented in parallel across branches can interact in complex ways, particularly when they affect related elements (e.g., method signatures, inheritance, and interface implementations), leading to overlapping changes that hinder automatic conflict resolution and increase integration effort.}

Our investigation reveals that, in our dataset, most refactoring types are associated with higher-effort merges, whereas merges without refactorings show lower rates of high-effort cases. Notably, \textit{Rename Attribute} is linked to a 400\% increase in merge effort, while structural refactorings such as \textit{Move Class} reach even stronger associations (+431\%). Other transformations, including \textit{Extract Variable} (+367\%) and \textit{Change Return Type} (+320\%), follow similar patterns, and even less frequent operations, such as \textit{Split Parameter}, show substantial associations with merge effort. We also find that both refactoring diversity and volume are independently associated with merge effort, affecting not only its occurrence but also its intensity. Finally, the co-occurrence of refactoring types across parallel branches shows stronger associations with merge effort than isolated transformations, particularly when combining structural changes with modifications to method signatures or data representations, whereas more localized changes appear less frequently among the most impactful combinations.

Our findings suggest that developers should avoid the concurrent application of structural and high-risk refactoring combinations across branches, favoring early integration, smaller increments, and short-lived branches. Monitoring refactoring volume and diversity can help anticipate integration risks, while tool support (e.g., CI bots, GitHub Actions, and Gerrit hooks) can proactively flag risky scenarios and guide merge timing decisions. Section~\ref{sec:resultsDiscussion} discusses these implications.

The remainder of this paper is organized as follows. Section~\ref{sec:methodology} describes the research process, dataset, and techniques. Section~\ref{sec:results} presents the results and answers the research questions. Section~\ref{sec:threats} outlines the threats to validity. Section~\ref{sec:related} reviews related work. Section~\ref{sec:conclusions} summarizes our contributions and future research directions.

\vspace{-2mm}
\section{Materials and Methods} \label{sec:methodology}

\begin{figure}[!htbp]
\centering
\includegraphics[width=3.4in]{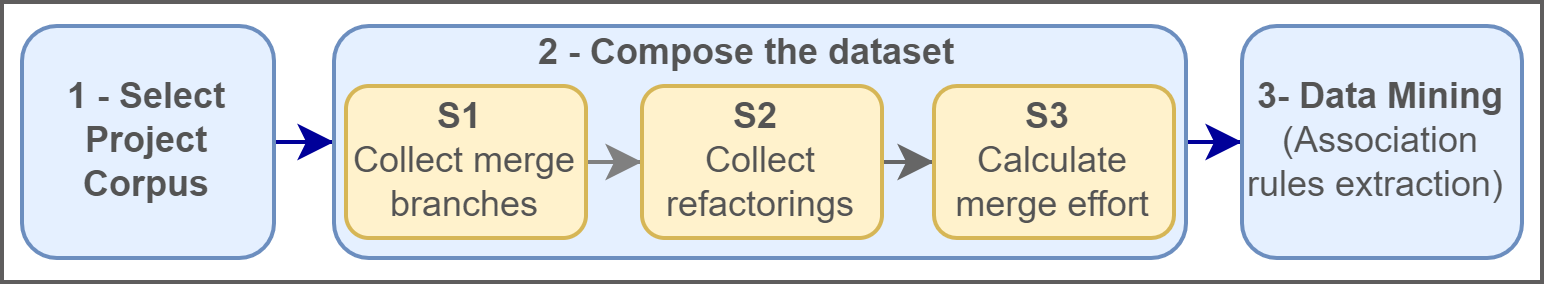}
\caption{The phases of the research process.}
\label{fig:phases_steps}
\end{figure}

Our research process comprises three phases (Figure~\ref{fig:phases_steps}). Phase one defines the project selection criteria (Section~\ref{sec:project_corpus}). Phase two collects refactoring and merge data and computes merge effort measures, composing the dataset (Section~\ref{sec:composet_dataset}). Phase three (Section~\ref{sec:association_rules}) applies association rule mining~\cite{han2022} to address the research questions (Section~\ref{sec:introduction}), uncovering patterns that may be overlooked by manual analysis. The data, scripts, and reproducibility guidelines are available in the replication package\footnote{\url{https://github.com/gems-uff/refactoring-merge}}. Data extraction and mining were conducted between May and July 2025. The dataset spans from August 2008 to July 2025, with an average project history of 12.8 years (range: 4.62–16.90 years).

\vspace{-2mm}
\vspace{-2mm}

\subsection{\textbf{Select Project Corpus}} 
\label{sec:project_corpus}

We constructed the corpus from mature, relevant open-source GitHub projects using the GraphQL API (v4)\footnote{\url{https://docs.github.com/pt/graphql}}. We identified public repositories that met these criteria: (i) not forks, (ii) $\ge$5,000 stars, (iii) not archived, and (iv) received at least one push in the last three months. Following Kalliamvakou et al.~\cite{kalliamvakou2014}, we excluded forks to ensure single project representation. We verified that no forked repositories met the selection criteria. The 5,000-star threshold ensured corpus relevance and widely recognized repositories~\cite{borges2018}. Excluding archived and inactive repositories ensured ongoing development. The search resulted in a final set of 6,161 repositories.

After initial selection, we further filtered the 6,161 repositories by analyzing the contributors and the commits metadata. We retained repositories with (i) $\ge$10 contributors and (ii) $\ge$3,000 commits in the default branch. The contributor filter excluded personal or coursework projects, ensuring collaborative, production-level software~\cite{kalliamvakou2014}. The commit threshold eliminated immature or short-term projects, focusing on sustained development. The contributor filter yielded 5,714 repositories. Of these, 1,989 met the $\ge$3,000 commit criterion.

From the 1,989 repositories, we selected only Java-primary proj\-ects, as reported by GitHub. We focused on Java due to several factors: (i) Java's popularity (TIOBE index~\cite{tiobe2024}, Stack Overflow Developer Survey~\cite{stackoverflow2024}). Although Java is not the most popular language overall, its consistent ranking among the top positions demonstrates its enduring relevance and influence within the software development community; (ii) the need to ensure consistency and comparability with our prior work~\cite{oliveira2023}, which was also conducted on Java systems; (iii) the reduced confounding effects associated with language-specific characteristics, such as differences in syntax, refactoring practices, and development ecosystems; and (iv) the suitability of Java for refactoring analysis, given its strong object-oriented nature, mature tooling support, and large availability of long-lived and well-maintained open-source repositories. These filters yielded 112 repositories.

We manually inspected the 112 repositories via GitHub pages and, when available, their official websites. The inspection removed non-software projects or those with non-English documentation. One repository contained only documentation (e.g., books, manuals), explaining its classification; we removed it, leaving 111 projects. We focused on English documentation for complete artifact understanding. We excluded seven repositories with primarily Chinese documentation, as inconsistent translation quality risked the accuracy and reliability of the analysis. The exclusions resulted in 104 repositories. 

Our study considers merge commits resulting from standard \texttt{git merge} operations that integrate \textit{named branches}, including those associated with \textit{pull request} integrations, as well as merges created within pull request branches prior to integration. We identify a commit as a merge commit if it has exactly two parents in the Git history. Additionally, we considered only non-fast-forward merge commits as valid, since we aimed to assess merging effort. A fast-forward merge moves the target branch pointer to the source branch without creating a new merge commit, provided there is a direct linear relationship between these two branches. In contrast, \texttt{git merge -\--no-ff} performs a non-fast-forward merge, creating a new merge commit and preserving history and branch structure even when a fast-forward merge is possible. We calculated the distribution of valid merge commits per project. Using Tukey’s fences~\cite{barnett1994}, defined by the lower bound ($Q1 - 1.5 \times IQR$) and upper bound ($Q3 + 1.5 \times IQR$), we analyzed the distribution of valid merge commits to establish selection criteria and filter out extreme cases.

To ensure a homogeneous dataset, we excluded projects based on the distribution of valid merge commits: 13 projects exceeding the upper threshold (5,503 merges) and 27 projects below the first quartile (218 merges) were removed. The final corpus comprises 64 systems, a scale consistent with or exceeding related studies (e.g., 10--70 projects)~\cite{bibiano2024, vale2024, ferreira2025}. We further discarded 28.3\% of \texttt{--no-ff} merges and excluded octopus merges (more than 2 parents), which are inherently conflict-free. This resulted in 91,270 merge commits.

These criteria filter trivial cases without biasing the dataset toward high-effort outcomes. By not constraining projects based on refactoring frequency, the corpus reflects a natural distribution of real-world development practices. This allows for an unbiased assessment of refactoring influence across diverse merge scenarios. Detailed project-level information is available in our replication repository\footnotemark[2].

\vspace{-2mm}
\subsection{\textbf{Compose the Dataset}}
\label{sec:composet_dataset}

As illustrated in Figure~\ref{fig:phases_steps}, the second phase was carried out in three steps, each implemented in a separate Python script: (S1)~identifying the commits that comprise each branch of the merge commits, (S2)~collecting the refactorings implemented within the commits of each branch, and (S3)~calculating the merge effort.

\begin{figure}[!htbp]
\centering
\includegraphics[width=3.2in]{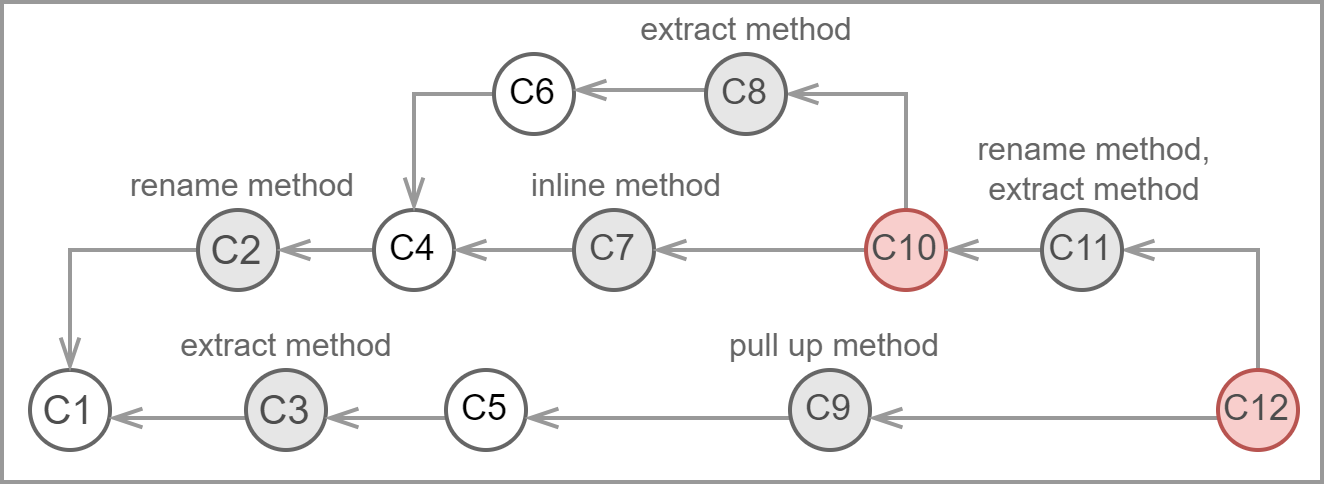}
\caption{Implementing refactorings on merge commit branches.}
\label{fig:dag}
\end{figure}

\href{https://github.com/gems-uff/refactoring-merge/blob/master/src/script_1_colllect_branches.py}{Script S1} collected all commits from the branches linked to the merge commits found in the repositories. Figure~\ref{fig:dag} shows an example of version history with two merge commits (C10, C12) highlighted in red. For C10, the script identifies branch\_1$_{C10}$ contains C7, while branch\_2$_{C10}$ includes C6 and C8. Since Script S1 handles nested merges, for C12, it identifies branch\_1$_{C12}$ contains C3, C5, C9. Branch\_2$_{C12}$ includes C2, C4, C6, C7, C8, C11, encompassing all commits from C10's branches. We used Python's pygit2\footnote{\url{https://www.pygit2.org/}} to gather project versioning data.

\vspace{-2mm}
\vspace{-2mm}

\href{https://github.com/gems-uff/refactoring-merge/blob/master/src/script_2_collect_refactorings.py}{Script S2} identified refactorings in merge commit branches. Figure~\ref{fig:dag} shows commits that contain refactorings highlighted in gray. To detect refactorings, we used \textit{RefactoringMiner} (v3.1, March 2026), which can detect more than 100 refactoring types. Its authors reported 99.6\% precision and 99.4\% recall, establishing it as state-of-the-art for automated refactoring detection~\cite{alikhanifard2025}. After script execution, we computed the number of refactorings by type in merge commit branches. We analyzed 700,308 commits using RefactoringMiner.

In this study, refactorings are identified in the branches' development history prior to the merge commit. Thus, our analysis focuses on refactoring activities introduced before the integration step and excludes refactorings performed during merge resolution. We examines 33 types of refactorings, including 26 from Fowler's catalog~\cite{fowler2018}: \textit{Change Return Type}, \textit{Extract Attribute}, \textit{Extract Class}, \textit{Extract Interface}, \textit{Extract Method}, \textit{Extract Subclass}, \textit{Extract Superclass}, \textit{Extract Variable}, \textit{Inline Method}, \textit{Inline Variable}, \textit{Merge Attribute}, \textit{Move Attribute}, \textit{Move Class}, \textit{Move Method}, \textit{Pull Up Attribute}, \textit{Pull Up Method}, \textit{Push Down Attribute}, \textit{Push Down Method}, \textit{Rename Attribute}, \textit{Rename Class}, \textit{Rename Method}, \textit{Rename Parameter}, \textit{Rename Variable}, \textit{Split Attribute}, \textit{Split Parameter}, and \textit{Split Variable}. We also included seven additional types defined by Tsantalis et al.\cite{tsantalis2020}: \textit{Change Parameter Type}, \textit{Change Variable Type}, \textit{Merge Parameter}, \textit{Merge Variable}, \textit{Parameterize Variable}, \textit{Replace Attribute}, and \textit{Replace Variable with Attribute}. We selected these refactoring subsets based on recent findings indicating their frequent use by developers in practice and their exploration in recent research~\cite{bibiano2021, liu2025, wang2025, alameer2026}.

Finally, \href{https://github.com/gems-uff/refactoring-merge/blob/master/src/script_3_calculate_merge_effort.py}{Script S3} calculated the merge effort. We investigated the relationship between refactorings and this effort using a metric defined by Prud\^encio et al.\cite{prudencio2012} and implemented by Moura and Murta~\cite{moura2018}. We computed the merge effort by calculating merge code churn via a \texttt{diff} between the base (common ancestor) and merged versions. This generates a multiset including all committed actions. We subtracted the branch-level multiset from the merge multiset. The resulting multiset contains only lines that were added or explicitly removed during the merge, and the merge effort corresponds to the total number of actions taken. The complete formulation and implementation details are available in our replication package\footnotemark[2].

Conflicts are categorized into: (i) textual, (ii) build (syntactic/static semantic), and (iii) test (behavioral) conflicts \cite{silva2024}. By defining merge effort as the lines added or removed during integration, our approach accounts for all three types. Unlike Git, which only detects textual conflicts, our metric captures the manual edits required to resolve build and test inconsistencies. Consequently, we avoid underestimating the actual integration effort, a common limitation in studies focused solely on textual conflicts.

We define merge effort as the effort required to reconcile concurrent changes during code integration, which differs from the effort required to perform refactorings. Our analysis focuses on how refactorings introduced in parallel branches interact during integration. Rather than measuring the effort to apply refactorings, we examine how they interfere with independent edits in another branch. For example, while a \textit{Rename Attribute} may require updating multiple references, we do not measure the effort to apply such transformations, but rather how they interfere with concurrent edits in another branch. Refactorings that propagate across multiple references or alter structural relationships may overlap with concurrent changes. In such cases, merge tools based primarily on textual comparison may struggle to automatically integrate these edits, requiring developers to interpret the refactoring intent and reconcile changes across different code locations. This helps explain why certain refactoring types are more frequently associated with merge effort in our dataset.

Building on this definition, we operationalize merge effort using code churn, measured as the number of added and removed lines in merge commits. This choice enables automated large-scale analysis across thousands of merge scenarios and aligns with prior work that adopts code churn as a proxy for maintenance effort and change complexity~\cite{sjoberg2012, olsson2017, carka2022, jesse2023, bessghaier2025}. Our metric captures the additional changes required to reconcile concurrent modifications during merge resolution, rather than the total changes introduced by refactorings, thereby focusing on the integration process. Consequently, large changes do not necessarily imply high merge effort, while small changes may still require substantial reconciliation. Furthermore, VCSs such as Git operate at the line level; merge conflicts are detected and resolved line by line, making line-based churn naturally aligned with the granularity at which merge effort manifests in practice. This notion is consistent with prior work~\cite{sjoberg2012}, which reported a moderate-to-strong correlation ($\rho = 0.59$–$0.66$) between code churn and maintenance effort. For example, when independent changes affect separate files or non-overlapping regions, merge effort is typically negligible. In contrast, integrating changes that involve structural modifications or overlapping edits may require substantial reconciliation effort.

While more fine-grained metrics, such as Levenshtein edit distance~\cite{yujian2007}, can capture change magnitude at a lower granularity~\cite{chowdhury2022}, they operate at a granularity that is not directly aligned with the underlying merge mechanism, and tend to be more computationally expensive, which has limited their adoption in large-scale studies~\cite{aguado2022}, and do not fully capture the reconciliation effort involved in merge resolution. Time-based measures are often unreliable in open-source contexts because accurately identifying the start and end of merge-resolution activities is not feasible from repository data alone. Qualitative analysis is inherently limited in scalability and would not be applicable to the 91,270 merge commits analyzed in this study. We also acknowledge that code churn does not capture all dimensions of merge effort, such as cognitive and coordination aspects or the time required for resolution.

To further assess the validity of code churn as a proxy for merge effort, we conducted a targeted manual validation on a stratified sample of 180 merge commits with non-zero effort (Effort$>$0), considering the three effort levels (\textit{low}, \textit{medium}, and \textit{high}), and the same distribution criteria in each group, as defined in RQ1. For each case, we inspected the combined \texttt{diff} (\textit{git diff merge\_sha1 merge\_sha1\^{}1 merge\_sha1\^{}2}) to verify whether the observed churn corresponded to manual reconciliation actions, with particular attention to textual conflicts involving refactorings across branches. Overall, we identified such conflicts in 39 out of the 180 analyzed merges, distributed as 3, 15, and 21 cases in the low-, medium-, and high-effort groups, respectively. 

These conflicts involved refactoring types such as \textit{Change Parameter Type} (22), \textit{Rename Method} (16), \textit{Change Return Type} (10), \textit{Rename Class} (7), \textit{Move Method} (4), \textit{Rename Attribute} (4), \textit{Extract Method} (2), \textit{Extract Class} (2), \textit{Move Class} (2), \textit{Extract Interface} (1), and \textit{Split Attribute} (1). Although this analysis is based on a sample, the presence and distribution of these refactoring types are consistent with the patterns observed in our large-scale analysis, particularly those associated with changes to method signatures, structural reorganization, and interface modifications. This alignment provides additional support for interpreting code churn as a meaningful proxy for merge effort in the context of concurrent refactorings. The examples identified during this inspection are available in our replication package\footnotemark[2].

\vspace{-2mm}
\subsection{\textbf{Data Mining - Association Rules Extraction}} \label{sec:association_rules}

We used association rule mining~\cite{han2022} to uncover relationships between refactoring types and merge effort. This exploratory technique identifies patterns within software repositories without predefined hypotheses, generating insights that can confirm or expand existing software engineering theories~\cite{han2022}. By identifying frequent associations among attributes, this approach reveals trends that might otherwise go unnoticed during manual analysis.

To investigate the relationship between refactoring and merge effort, we analyzed 33 refactoring types (represented as $refactoring\_type\_x$ for branches $b1$ and $b2$) alongside code churn as the effort metric. As summarized in Table~\ref{table:attributes}, our dataset comprises 67 attributes: 33 refactoring types per branch plus the merge effort indicator.

\vspace{-2mm}
\begin{table}[ht]
\centering
\caption{Attributes considered in the analysis.}
\label{table:attributes}
\scalebox{0.95}{
\begin{tabular}{ll}
\toprule[1.0pt]
\textbf{Attribute} & \textbf{Description} \\
\midrule
$refactoring\_type\_x_{b1}$ & \makecell[l]{number of refactoring type $x$ \\implemented in branch 1.} \\ \\
$refactoring\_type\_x_{b2}$ & \makecell[l]{number of refactoring type $x$ \\implemented in branch 2.} \\ \\
$effort$ & \makecell[l]{number of new lines included or \\ excluded in the merge commit \\ (code churn)} \\
\bottomrule[1.0pt]
\end{tabular}
}
\end{table}
\vspace{-2mm}

The method proposed in this work uses the concept of multidimensional association rules~\cite{han2022}. Given a relation (or table) $D$, a multidimensional association rule $X \rightarrow Y$, defined on $D$, is an implication of the form: $X_{1}\wedge X_{2}\wedge\cdots\wedge X_{n} \rightarrow Y_{1}\wedge Y_{2}\wedge\cdots\wedge Y_{m}$, where $n\geq 1, m\geq 1,$ and $X_{i} (1\leq i\leq n)$ as well as $Y_{j} (1\leq j\leq m)$ are conditions defined in terms of the distinct attributes of $D$~\cite{witten_data_2016, han2022}.

The rule $X\rightarrow Y$ indicates, with a certain assurance, that antecedent~$X$ is associated with consequent~$Y$. We evaluate rule relevance using three standard measures: \textit{Support}, \textit{Confidence}, and \textit{Lift}~\cite{witten_data_2016}. \textit{Support} captures how frequently both the antecedent and the consequent occur together in the dataset, while \textit{Confidence} reflects how often the consequent is observed when the antecedent occurs. These two measures are used to filter the extracted rules, and only rules that meet predefined minimum thresholds are retained for analysis.

For a straightforward illustration of \textit{Support} and \textit{Confidence} calculation, we refer to Table~\ref{table:basededadosexemplo}. This table presents three columns: i) line id, ii) number of \textit{Extract Method} implemented in branch 1; iii)  number of \textit{Rename Method} implemented in branch 2; iv) merge effort (as ``true''/``false''). Each row corresponds to a unique merge commit. We discretized refactorings into four levels: zero (``0''), units (``u''), dozens (``d''), and hundreds or more (``$\geq 100$''). For rule $R$: \textit{extract\_method\_b1~=~``d''}$~\wedge$ \textit{rename\_method\_b2~=~``d''} $ \rightarrow$ \textit{effort~=``true''}, we identify four records in $D$ meeting all three conditions (rows 1, 4, 6, 8). This results in $T_{(X \cup Y)}=4$. Given $D$ has 8 entries ($T=8$), we calculate \textit{Sup(R)} as 50\%~(4/8). For \textit{Confidence}, the rule achieves 80\%~(4/5) since antecedent conditions (\textit{extract\_method\_b1~=~``d''} $\wedge$ \textit{rename\_method\_b2~=~``d''}) hold in five entries ($T_{X}=5$): rows 1, 3, 4, 6, 8.

\begin{table}[ht]
\centering
\caption{Refactoring type and merge effort dataset sample.}
\label{table:basededadosexemplo}
\scalebox{0.95}{
\begin{tabular}{cccc}
\toprule[1.0pt]
\textbf{\#} & \textbf{extract\_method\_b1} & \textbf{rename\_method\_b2} & \textbf{effort} \\

\midrule
1 & ``d'' & ``d''         & ``true''  \\ 

2 & ``$\geq100$'' & ``$\geq100$'' &  ``false'' \\ 

3 & ``d'' & ``d''         &  ``false'' \\ 

4 & ``d'' & ``d''         &  ``true'' \\ 

5 & ``u'' & ``u''         & ``false'' \\ 

6 &  ``d'' &``d''         & ``true'' \\ 

7 & ``0'' & ``0''         & ``false'' \\ 

8 & ``d'' & ``d''         & ``true'' \\ 
\bottomrule[1.0pt]
\end{tabular}
}
\end{table}

We also consider \textit{Lift} to assess the strength of the association between antecedent and consequent. Intuitively, \textit{Lift} indicates how much more frequently the consequent occurs when the antecedent is present, compared to what would be expected if they were independent. A \textit{Lift} value equal to 1 indicates independence, whereas values greater than 1 indicate a positive dependence, and values lower than 1 indicate a negative dependence.

Using rule $R$ as an example, the consequent's \textit{Support} -- \textit{Sup(effort = ``true'')} -- in $D$ is 50\%, representing entries meeting \textit{effort = ``true''}. Rule $R$'s \textit{Lift} is 1.6, calculated as \textit{Lift(R) = 80/50 = 1.6} (80\% is rule \textit{Confidence}). This \textit{Lift} suggests that when dozens of \textit{Extract Method} are in branch 1 and \textit{Rename Method} in branch 2, the merge effort probability increases by 60\%. In other words, \textit{effort = ``true''} probability in $D$, initially 50\%, rises to 80\% given antecedent conditions (\textit{extract\_method\_b1 = ``d''} $\wedge$ \textit{rename\_method\_b2 = ``d''}).  

We applied the Apriori algorithm~\cite{agrawal_fast_1994} to extract association rules, using its implementation in a Python library\footnote{\url{http://rasbt.github.io/mlxtend/user_guide/frequent_patterns/apriori/}}. We selected Apriori as a well-established method for frequent pattern mining, which generates candidate itemsets and prunes the search space using the anti-monotonic property of support, ensuring that infrequent itemsets do not generate supersets. This guarantees completeness with respect to the specified support and confidence thresholds. Alternative algorithms (e.g., FP-Growth~\cite{han2000} and Eclat~\cite{zaki2000}) produce the same rules for a given dataset and thresholds, differing mainly in computational efficiency, depending on the dataset’s characteristics. As our dataset size and minimum support do not impose computational constraints, Apriori provides a simple, well-understood, and reproducible solution. We set a minimum support threshold of 20 merge commits to reduce coincidental or non-representative patterns.

For RQ3, to analyze refactoring types' influence in parallel merge branches, we extracted association rules. We identify combinations that frequently co-occur across branches and are linked to the merge effort. These rules show two specific refactorings, each in a separate branch, increasing merge effort chances (e.g., \textit{refac\_typeX\textsubscript{b1}=``true''}\linebreak~$\land$~\textit{refac\_typeY\textsubscript{b2}=``true''}~$\Rightarrow$~\textit{effort=``true"}). To identify if concurrently applied refactorings in different branches associate with increased merge effort, we adapted rule formulation to capture refactoring \textit{co-occurrence across branches} regardless of specific branch assignment (1 or 2). That is, we treated \textit{refac\_typeX} in branch 1 and \textit{refac\_typeY} in branch 2 as equivalent to the inverse. To abstract this, we redefined rule support to reflect the total merge commits where both refactorings occur across branches, independently of their implementation branch. We calculated this adjusted support by summing the occurrences of both variants and subtracting the overlap where they appeared in both branches. This aggregation allowed us to compute a single \textit{Lift}, measuring the frequency of merge effort when two refactorings occur in parallel in branches b$_\alpha$ and b$_\beta$ (e.g., \textit{Sup(rule)\textsubscript{new} = Sup(refac\_typeX\_b$_{\alpha}$=``true''}~$\land$~\textit{refac\_typeY\_b$_\beta$=``true''}\linebreak$\Rightarrow$ \textit{effort=``true'')}). We applied the same logic to compute\linebreak antecedent support, \textit{{Sup(ant)$_{\text{new}}$ = Sup(refac\_typeX\_b$_{\alpha}$=``true"~$\land$ refac\_typeY\_b$_\beta$=``true"})}, then calculated adjusted \textit{Lift} as: \textit{Lift(rule)$_{\text{new}}$~\linebreak= Sup(rule)$_{\text{new}}$ \,/\,(Sup(ant)$_{\textit{new}}$ $\times$ Sup(effort=``true"))}. 

For completeness, the formal definitions of \textit{Support}, \textit{Confidence}, and \textit{Lift} are available in our replication package\footnotemark[2].

\vspace{-2mm}
\section{Results and Discussion}
\label{sec:results}

This section presents and analyzes the results, addressing the research questions introduced in Section~\ref{sec:introduction}.

\vspace{-2mm}
\subsection{\textbf{RQ1}: Which Refactoring Types Most Influence the Occurrence of Merge Effort?} \label{sec:resultsRQ1}

\begin{figure}[!t]
\centering
\includegraphics[width=3.38in]{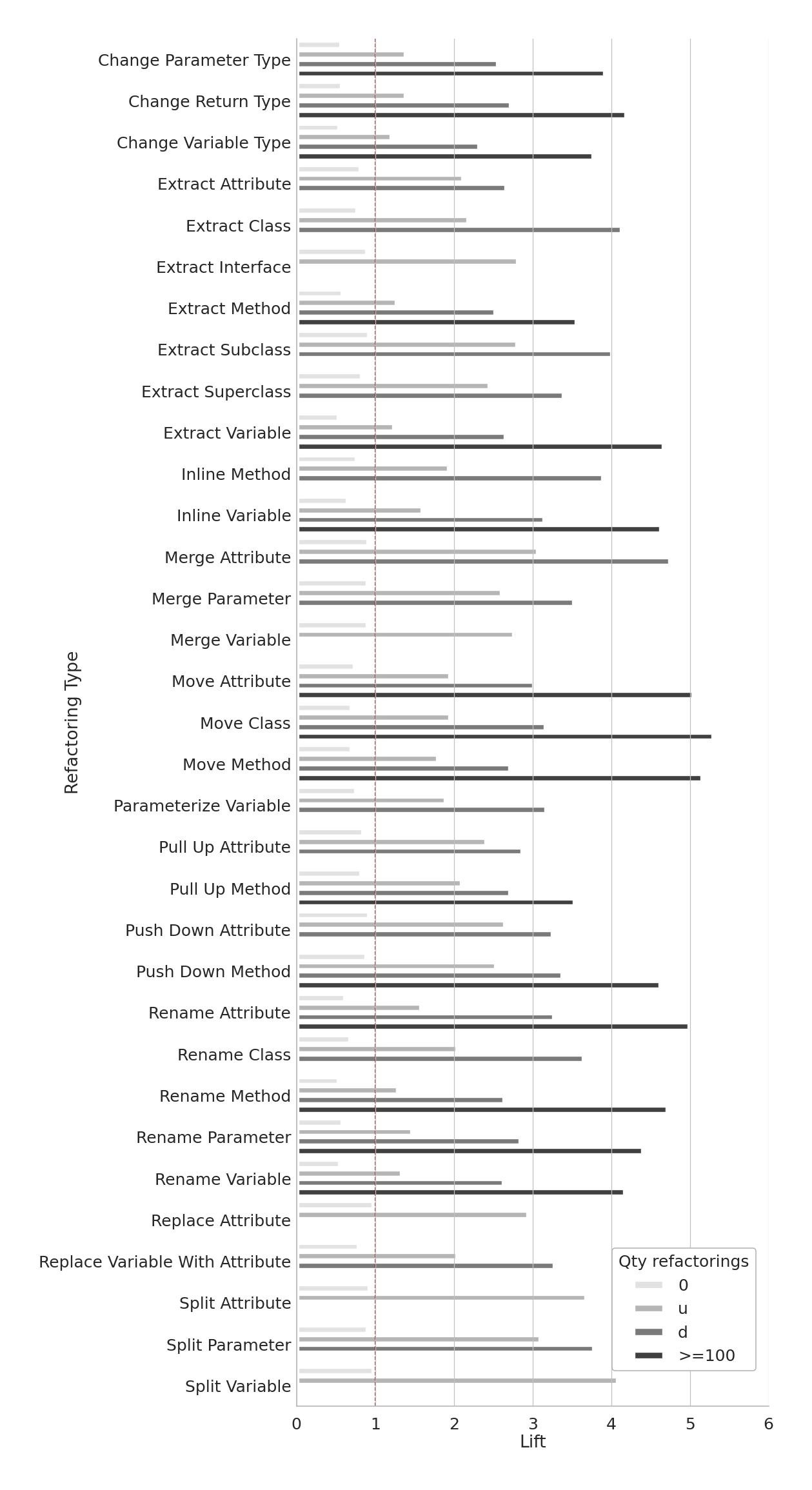}
\vspace{-15pt}
\caption{Influence of each refactoring type on the occurrence of merge effort (effort = ``true'')}
\label{fig:rq1_chart}
\end{figure}

This RQ investigates how each refactoring type's frequency influences merge effort, independent of the branch where it was applied. Figure~\ref{fig:rq1_chart} summarizes results. The y-axis displays the 33 analyzed refactoring types from our study, listed in alphabetical order. Initially, we discretized merge effort into a binary variable and extracted association rules for each refactoring type based on refactoring counts, categorized into four ranges: \textit{qty\_refac\_type=``0''~(zero) / ``u''~(units, 1--9) / ``d''~(dozens, 10--99) / ``$\geq100$''~(hundreds or more)} $\rightarrow$ \textit{effort = ``true''}. We used a grayscale gradient to distinguish these ranges: lightest gray for \textit{zero} refactorings and darkest gray for \textit{hundreds or more}. The x-axis shows each extracted rule's \textit{Lift} value, indicating the strength of the relationship between the antecedent (refactoring frequency) and the consequent (merge effort occurrence). We set the merge effort to ``true'' as we focused on analyzing which refactorings tend to result in non-zero merge effort.

\setlength{\textfloatsep}{5pt} 

Our previous work~\cite{oliveira2023} shows that the number of refactorings is associated with merge effort. Figure~\ref{fig:rq1_chart} confirms that this pattern persists for individual refactoring types. For all types, refactoring absence (``0'' -- lighter gray bar) shows a lower frequency of merge effort occurrences (\textit{Lift} $<$ 1). Intuitively, \textit{Lift} values greater than 1 indicate that merge effort occurs more frequently than expected given the baseline distribution, while values lower than 1 indicate a lower-than-expected occurrence. Thus, a Lift of 3.00 means that the merge effort occurs approximately three times more frequently in the presence of the antecedent than expected under independence. 

For instance, when the \textit{Extract Method} refactoring is absent in both merge branches (\textit{Extract Method = ``0''} $\rightarrow$ \textit{effort = ``true''}) yields a \textit{Lift} of 0.59, meaning that merge effort occurs 41\% less frequently. In contrast, merges that involve the \textit{Extract Method} refactoring show higher frequencies of merge effort, even in small quantities: ``u'' -- 28\% increase (\textit{Lift} = 1.28), ``d'' -- 154\% increase (\textit{Lift} = 2.54), and ``$\geq100$'' -- 256\% increase (\textit{Lift} = 3.56). All corresponding graph bars show \textit{Lift} values greater than 1. In nearly all scenarios, except \textit{Extract Attribute}, \textit{Lift} consistently increases as the number of implemented refactoring increases. For instance, implementing a large number of \textit{Rename Attribute} refactorings (``$\geq100$'') resulted in a \textit{Lift} apex of 5.00, indicating that merge effort occurs 400\% more frequently.

For certain types of refactoring, the graph reveals that we did not identify rules that meet the minimum support threshold (20 merge commits) for all discretization ranges. The absence of bars for types such as \textit{Split} refactorings (the last three on the x-axis), which separate attributes, parameters, and variables, reflects their low overall frequency. This outcome arises as some refactoring types are inherently less frequent. A significant \textit{Lift} increase occurs for the three \textit{Split} types when increasing from zero (``0'') to units (``u''). Results show that implementing even a few refactorings of these rarer types can significantly increase the chances of having a merge effort. Notably, among the three \textit{Split} refactorings, \textit{Split Parameter} appears more frequently, also meeting the minimum support threshold for the ``d'' range, where it exhibits a \textit{Lift} of 3.79 (+279\%), following the same increasing trend observed across discretization levels.

Building on this observation, several refactoring types exhibit a clear increasing pattern in the occurrence of merge effort as their frequency increases, forming a ``staircase'' effect across the discretization bands. Among the most critical cases, \textit{Move Class}, \textit{Move Method}, and \textit{Move Attribute} stand out, all reaching \textit{Lift} values above 5.00 in the ``$\geq\!100$'' range. For instance, \textit{Move Class} increases from 1.96 in ``u'' to 5.31 in ``$\geq\!100$'', while \textit{Move Method} rises from 1.80 to 5.16, and \textit{Move Attribute} from 1.96 to 5.02. Other refactorings also demonstrate strong growth patterns, such as \textit{Extract Variable} (from 1.25 to 4.67), \textit{Inline Variable} (from 1.61 to 4.64), and \textit{Push Down Method} (from 2.54 to 4.63). In addition, type-changing refactorings such as \textit{Change Return Type} and \textit{Change Parameter Type} show consistent increases, reaching 4.20 (+320\%) and 3.93 (+293\%), respectively. Renaming operations also follow this pattern, with \textit{Rename Attribute} showing one of the steepest escalations (from 1.59 to 5.00), and other renaming refactorings such as \textit{Rename Method}, \textit{Rename Parameter}, and \textit{Rename Variable} presenting increases above 200\% when comparing ``u'' to ``$\geq\!100$''. Overall, these results indicate that certain refactoring types not only correlate with merge effort, but that their impact intensifies substantially as their volume increases, highlighting a compounding effect of structural changes during parallel development.

To provide a more fine-grained analysis of merge effort, we extended its representation beyond a binary formulation to four levels: \textit{none}, \textit{low}, \textit{medium}, and \textit{high}. To mitigate the impact of the highly imbalanced distribution, we calibrated the discretization thresholds to obtain comparable support values across non-zero categories, focusing on the support of the consequent, as \textit{Lift} is influenced by its marginal probability.  Specifically, we partitioned the range of effort $> 0$ into three equal-frequency levels: \textit{low} ($>0$ and $\leq 5$), \textit{medium} ($>5$ and $\leq 20$), and \textit{high} ($>20$), each representing approximately one third ($\approx$33.33\%) of the non-zero instances. This configuration produced similar support values (\textit{low}: 0.0222, \textit{medium}: 0.0237, \textit{high}: 0.0200), reducing the influence of consequent support on \textit{Lift} and enabling more balanced comparisons. We then selected the five refactoring types most strongly associated with refactoring quantity and merge effort for analysis: \textit{Change Return Type}, \textit{Extract Variable}, \textit{Move Class}, \textit{Rename Attribute}, and \textit{Split Parameter}.

Focusing on \textit{Change Return Type}, we observed a clear and consistent progression across both refactoring quantity and effort levels. For \textit{low} effort, the \textit{Lift} increases from 0.73 to 1.33, 2.02, and 2.16 across ``0'', ``u'', ``d'', and ``$\geq100$'', respectively, corresponding to incremental increases of +82.19\% (``0'' to ``u''), +51.88\% (``u'' to ``d''), and +6.93\% (``d'' to  ``$\geq100$''). For \textit{medium} effort, the progression becomes more pronounced, with \textit{Lift} values rising from 0.56 to 1.52, 2.62, and 3.49, resulting in increases of +171.43\%, +72.37\%, and +33.21\%, respectively. The strongest escalation appears for \textit{high} effort, where \textit{Lift} increases from 0.44 to 1.30, 3.66, and 7.31, corresponding to +195.45\%, +181.54\%, and +99.73\%. These results indicate that, although the absence of this refactoring shows a negative association with merge effort (Lift $<$ 1), increasing quantities strongly shift the association toward higher effort levels, particularly for \textit{high} effort.

The remaining refactoring types exhibit similar progression patterns across effort levels, although they vary in magnitude and, in some cases, lack sufficient support for higher ranges. For \textit{Extract Variable}, the incremental increases from ``0'' to ``u'', ``u'' to ``d'' and ``d'' to ``$\geq100$'' are +98.46\%, +55.04\%, and +14.00\% for \textit{low} effort; +151.92\%, +97.71\%, and +65.25\% for \textit{medium}; and +153.33\%, +205.26\%, and +124.14\% for \textit{high}, indicating a consistent upward trend across all effort levels. 

For \textit{Move Class}, we observed increases of +114.81\% and +28.16\% for \textit{low} effort (``0'' to ``u'' and ``u'' to ``d''), although no rule meets the minimum support threshold for $\geq100$ in this level; for \textit{medium}, the progression continues with +200.00\%, +21.60\% and +84.17\%; and for \textit{high}, with +238.98\%, +146.50\%, and +103.65\%. A similar pattern appears for \textit{Rename Attribute}, with increases of +105.41\% and +41.45\% for \textit{low} effort (with no support for $\geq100$); +170.49\%, +90.91\%, and +33.97\% for \textit{medium}; and +220.00\%, +192.50\%, and +105.56\% for \textit{high}. Finally, \textit{Split Parameter}, despite its lower frequency, follows the same overall trend whenever sufficient support is available, with increases of +130.53\% for \textit{low}, +192.47\% for \textit{medium}, and +445.24\% for \textit{high} in the transition from ``0'' to ``u''; however, subsequent ranges (``u'' to ``d'' and ``d'' to ``$\geq100$'') do not meet the minimum support threshold. Overall, these results indicate that the incremental growth of association strength remains consistent across different refactoring types, even in the presence of data sparsity for less frequent transformations.

To illustrate in practice the merge effort associated with certain types of refactorings, we conducted a manual inspection of the combined diff of a merge commit (\texttt{bcaf1c}) from the \textit{shardingSphere} project, focusing on the \textit{Move Class} refactoring type. In our results, \textit{Move Class} showed a strong association with the occurrence of merge effort, as indicated by increasing \textit{Lift} values with higher numbers of such refactorings across the merge branches. In this commit, RefactoringMiner detected 272 \textit{Move Class} refactorings, while the merge effort amounted to 31 lines added or removed. From the combined diff, we identified 19 added import statements, a subset of which is illustrated in Code~\ref{cod:MergeEffortMoveClass}. Manual inspection of the source code confirmed that these changes correspond to 14 distinct classes that were moved from the same package as their callers to different packages, requiring explicit import declarations. This provides concrete evidence that structural refactorings, such as \textit{Move Class}, directly contribute to merge effort through dependency adjustments. 

\begin{listing}[htbp]
\caption{Example of merge effort associated with \textit{Move Class} refactorings in commit \texttt{bcaf1c} of the \textit{shardingSphere} repository.}
\label{cod:MergeEffortMoveClass}

\begin{lstlisting}[
    basicstyle=\ttfamily\fontsize{7}{5.8}\selectfont,
    escapeinside=@@,
    columns=fullflexible,
    keepspaces=true,
    aboveskip=0pt,
    belowskip=0pt
]
@\colorbox{diffadd}{\makebox[\linewidth][l]{\ttfamily\fontsize{7}{5.8}\selectfont ++ import org.apache(...).SQLStatement;}}@
@\colorbox{diffadd}{\makebox[\linewidth][l]{\ttfamily\fontsize{7}{5.8}\selectfont ++ import org.apache(...).expression.SQLExpression;}}@
\end{lstlisting}
\end{listing}

\MyBox{\textbf{Answer to RQ1}: Refactoring types exhibit a ``staircase'' pattern: higher frequencies consistently correlate with increased merge effort. While the absence of refactorings shows low association (\textit{Lift} $<$ 1), increasing volumes yield progressively higher \textit{Lift} values across \textit{low}, \textit{medium}, and \textit{high} effort levels. Structural refactorings show the strongest associations: \textit{Move Class} reached \textit{Lift} 5.31 (+431\%), followed by \textit{Move Method} and \textit{Move Attribute}. Renaming operations also escalate steeply, notably \textit{Rename Attribute} (\textit{Lift} 5.00; +400\%). Local transformations, such as \textit{Extract Variable} (\textit{Lift} 4.67; +367\%) and \textit{Inline Variable} (\textit{Lift} 4.64; +364\%), demonstrate similar growth. Type-changes (Eg. \textit{Change Return Type}) reached \textit{Lift} 4.20 (+320\%), while even infrequent operations like \textit{Split Parameter} showed substantial impact (\textit{Lift} 3.79; +279\%). These results suggest a compounding effect during parallel development, in which both refactoring type and volume can intensify the occurrence and magnitude of merge effort.}

\subsection{\textbf{RQ2}: Does the amount of different refactoring types influence the merge effort?} \label{sec:resultsRQ2}

\begin{figure}[!htbp]
\centering
\includegraphics[width=3.35in]{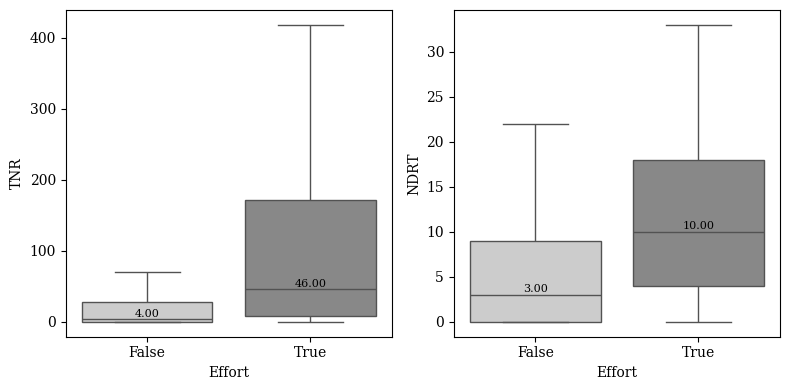}
\vspace{-10pt}
\caption{Distribution of the TNR and NDRT attributes considering the occurrence of merge effort (\textit{true} or \textit{false}).}
\label{fig:rq2_boxplot}
\end{figure}

This research question examines whether the variety of refactoring types is associated with a higher frequency of merge effort, regardless of the total number of refactorings. Because the number and diversity of refactorings often correlate, it is essential to disentangle their individual effects to determine whether diversity offers explanatory power beyond refactoring volume. We first explored the relationship between the total number of refactorings (TNR), the number of different refactoring types (NDRT), and merge effort (discretized as \textit{true} or \textit{false}), using boxplots shown in Figure~\ref{fig:rq2_boxplot}. The plot shows that both attributes present higher values when the effort is \textit{true}. Specifically, the median TNR rises from 4 to 46, and the median NDRT increases from 3 to 10. These raw changes correspond to 1,050\% and 233\% increases, respectively. However, due to the differing scales of the attributes (TNR: 0 -- 418; NDRT: 0 -- 33), proportional shifts relative to range provide more balanced insight -- approximately 10\% for TNR and 21\% for NDRT. This normalization suggests that NDRT shows a relatively more pronounced shift when adjusted for scale, despite TNR's larger absolute growth.

Since we observed that TNR $>0$ consistently implied NDRT $>0$, we hypothesized a strong association between these variables. To test this, we computed Spearman's~\cite{spearman1904} rank correlation, which confirmed a strong positive monotonic relationship ($\rho = 0.9771$, $p < 0.0001$). However, we also needed to verify whether these attributes provided non-redundant information. For this purpose, we calculated the Variance Inflation Factor (VIF)~\cite{obrien2007}, which measures the extent of multicollinearity among predictors. The resulting VIF of 1.31, well below the commonly accepted threshold of 5~\cite{obrien2007}, indicates low collinearity, supporting the interpretation that TNR and NDRT can be considered independently in subsequent analyses.

\begin{figure}[htbp]
\centering
\includegraphics[width=3.5in]{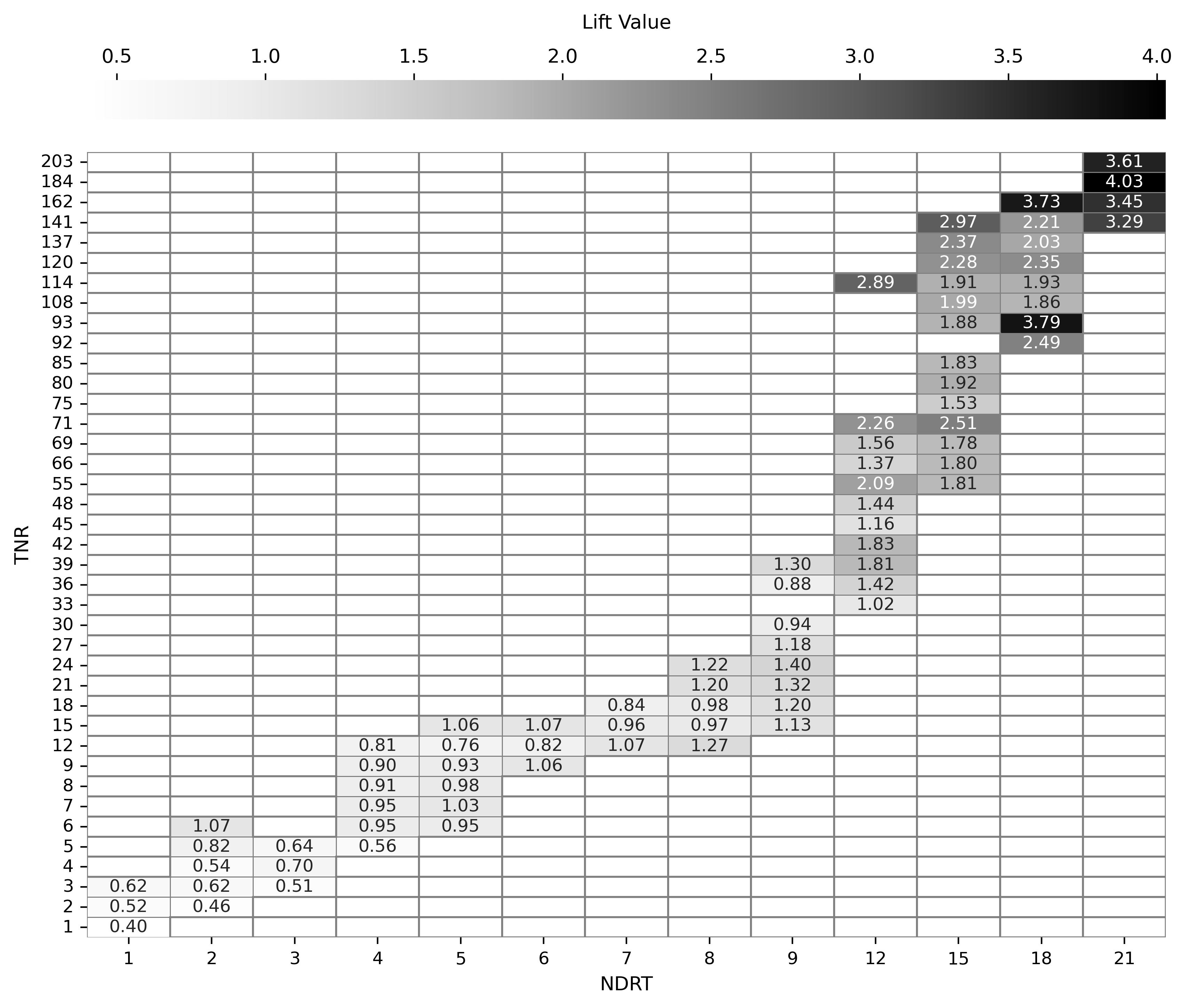}
\vspace{-15pt}
\caption{Heatmap showing the \textit{Lift} values of the extracted association rules:~\textit{TNR = ``value\textsubscript{i}'' and NDRT = ``value\textsubscript{j}'' $\rightarrow$ effort = ``true''}, where \textit{i} represents the row and \textit{j} the column.}
\label{fig:rq2_heatmap}
\end{figure}

Given the strong correlation between TNR and NDRT, we applied the Apriori algorithm~\cite{agrawal_fast_1994} to analyze whether each attribute, individually or in combination, is associated with the occurrence of merge effort. We collected association rules where the antecedent combines TNR and NDRT values, and the consequent represents the occurrence of merge effort. To visualize results, we constructed a heatmap (Figure~\ref{fig:rq2_heatmap}). Each cell shows the \textit{Lift} value of a rule satisfying the 20-tuple minimum support threshold. Specifically, we collected rules in the format:~\textit{TNR = ``i'' $\land$ NDRT = ``j'' $\rightarrow$ effort = ``true''}, where \textit{i} represents TNR (row) and \textit{j} NDRT (column). Empty cells indicate combinations yielding no rules due to failing the minimum support threshold. A practical example occurs when TNR = 93 and NDRT = 18 (rule:~\textit{TNR = 93 $\land$ NDRT = 18 $\rightarrow$ effort = ``true''}), resulting in a \textit{Lift} of 3.79 (+279\%).

To assess the individual influence of TNR and NDRT on merge effort, we analyzed percentage variations based on heatmap \textit{Lift} values. For each fixed TNR (rows), we performed adjacent pairwise comparisons by varying NDRT (columns). Then, we repeated this by fixing NDRT (columns) and varying TNR (rows). For each pair $(c1, c2)$, where $c1$ is the preceding cell and $c2$ the subsequent one within the same row or column (both containing the Lift value), we calculated the percentage difference using the formula:~\textit{(Lift\textsubscript{c2} - Lift\textsubscript{c1}) / Lift\textsubscript{c1}}. This metric quantifies the relative increase or decrease between consecutive \textit{Lift} values. For instance, fixing \textit{NDRT} = 2 and varying \textit{TNR} yields four pairwise percentage differences: between 1.07 and 0.82 = $(1.07 - 0.82)/0.82 = 30.49\%$; between 0.82 and 0.54 = $(0.82 - 0.54)/0.54 = 51.85\%$; between 0.54 and 0.62 = $(0.54 - 0.62)/0.62 = -12.90\%$; and between 0.62 and 0.46 = $(0.62 - 0.46)/0.46 = 34.78\%$.

We excluded comparisons involving isolated values or gaps between valid entries to avoid misleading variations. For example, fixing TNR = 6 and varying NDRT, we excluded 1.07 (TNR = 6 and NDRT = 2) due to a gap before 0.95 (NDRT = 4). Similarly, we excluded 2.89 in column-wise comparisons when a gap existed after 2.26 (TNR = 71 and NDRT = 12). After filtering, we computed the mean and standard deviation of the resulting percentage differences. TNR comparisons yielded a mean of 9.28\% and a standard deviation of 0.259. NDRT-related comparisons showed a mean of 7.88\% and a standard deviation of 0.258. Although modest, the results suggest a slightly stronger association for TNR, while indicating that both attributes contribute to the merge effort.

To complement the previous analysis and address merge effort intensity, we analyzed association rules with consequents representing effort levels (\textit{low}, \textit{medium}, \textit{high}), using the same discretization as in RQ1. We focused on representative NDRT groups (\textit{little}, \textit{some}, and \textit{much}), ensuring minimum support across all effort levels. For the \textit{little} diversity group: when NDRT = 2, all associations present \textit{Lift} values below 1: 0.69 (-31\%), 0.59 (-41\%), 0.40 (-60\%); indicating consistently lower frequencies of merge effort. A similar but weaker pattern is observed for NDRT = 4: 0.99 (-1\%), 0.97 (-3\%), 0.65 (-35\%); where \textit{low} and \textit{medium} remain near independence, but \textit{high} effort is still reduced ($\approx$35\%, \textit{Lift}=0.65). As diversity increases, associations become consistently positive. For \textit{some} group: when NDRT = 12, \textit{Lift} values are 1.55 (+55\%), 1.82 (+82\%), and 1.36 (+36\%), increasing to 1.61 (+61\%), 2.15 (+115\%), and 1.69 (+69\%) for NDRT = 14. This trend intensifies at higher levels in the \textit{much} diversity group: NDRT = 22 yielding 2.29 (+129\%), 3.20 (+220\%), and 3.99 (+299\%), and NDRT = 26 reaching 3.00 (+200\%), 2.51 (+151\%), and 5.81 (+481\%) for \textit{low}, \textit{medium}, and \textit{high}, respectively. This pattern becomes more pronounced at high levels of diversity. For instance, NDRT = 28, 29, and 31 are strongly associated with \textit{high} effort, with \textit{Lift} values of 6.67 (+567\%), 6.18 (+518\%), and 14.16 (+1,316\%), respectively. These cases indicate that merges involving many distinct refactoring types are substantially more likely to incur high merge effort. 

Overall, the results show that not only the volume but also the diversity of refactorings contribute to effort severity. While lower NDRT values are associated with reduced or near-independent effort levels, higher diversity consistently shifts the association toward more complex and effort-intensive merge scenarios, highlighting structural heterogeneity as a relevant risk factor.

\MyBox{\textbf{Answer to RQ2}: The number of different refactoring types (NDRT) and the total number of refactorings (TNR) both show significant and independent associations with merge effort. On average, each increase in NDRT corresponds to a 7.88\% rise in the frequency of merge effort occurrences, even when TNR remains constant. Similarly, a 9.28\% increase in refactoring volume (TNR) results in a proportional rise in merge effort frequency. These results highlight the independent and relevant contribution of diversity to the occurrence of merge effort and emphasize the importance of jointly considering both dimensions when evaluating or planning parallel code changes. Furthermore, this pattern is consistently observed across different levels of merge effort (\textit{low}, \textit{medium}, and \textit{high}), with higher refactoring diversity associated with progressively stronger associations with higher effort levels.}

\subsection{\textbf{RQ3}: How do combinations of refactoring types, applied concurrently in different branches, influence the merge effort?} \label{sec:resultsRQ3}

\begin{figure*}[!htbp]
\centering
\includegraphics[width=7.1in]{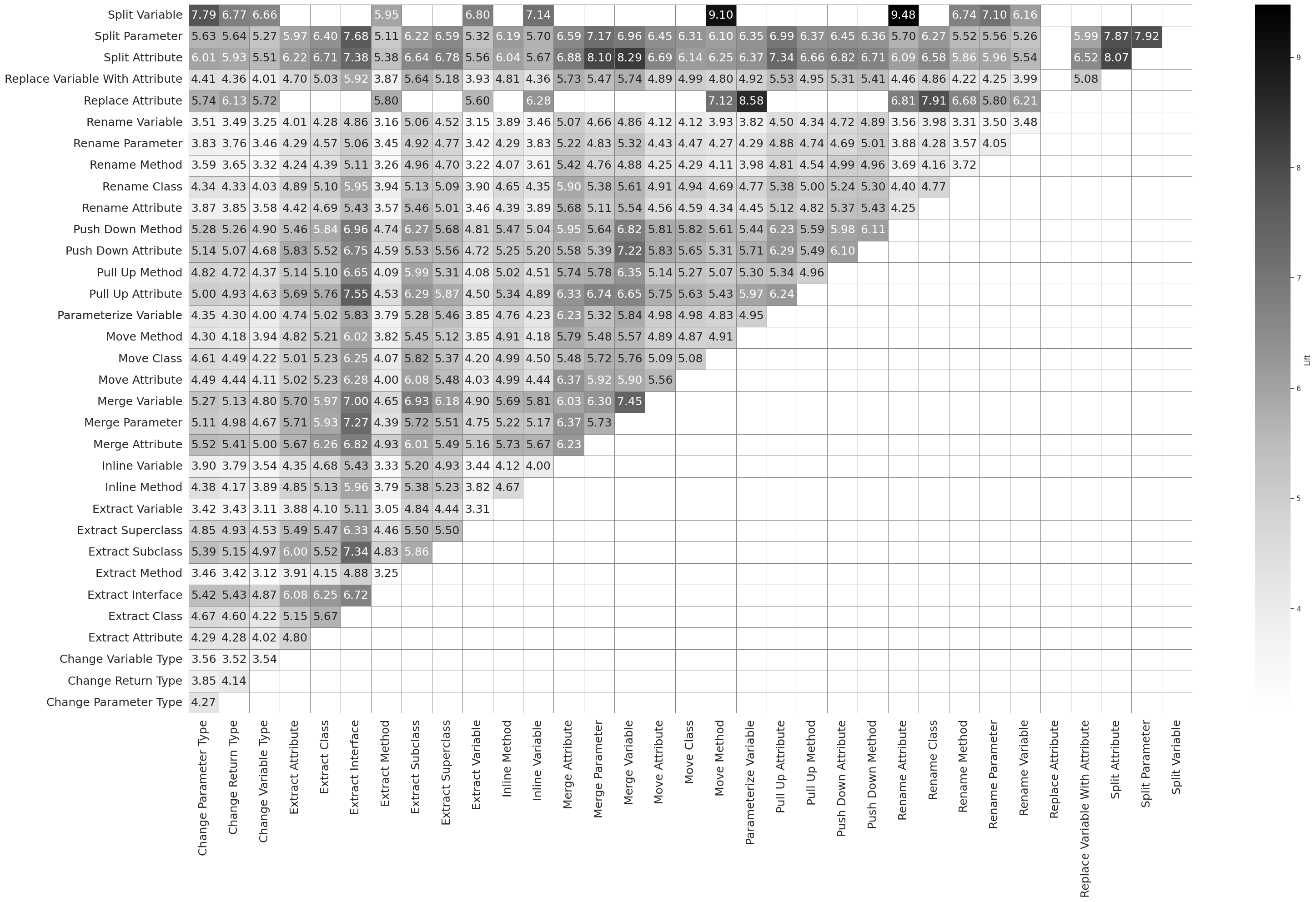}
\vspace{-5pt}
\caption{Heatmap of \textit{Lift} values associating refactorings introduced in parallel branches under the condition \textit{effort = true}.}
\label{fig:rq3_heatmap}
\end{figure*}

This RQ investigates whether specific combinations of refactoring types, applied concurrently in parallel branches, are associated with merge effort. Figure~\ref{fig:rq3_heatmap} presents a heatmap in which each cell reports the \textit{Lift} value of a refactoring pair under the condition \textit{effort = true}. We consider refactoring co-occurrence across branches irrespective of which branch each transformation was applied to. As described in Section~\ref{sec:association_rules}, \textit{Lift} values were recalculated under a symmetric formulation in which refactoring pairs are considered independently of the branch in which they occur. For instance, the rules \textit{Extract Method} in branch 1 and \textit{Extract Interface} in branch 2 $\Rightarrow$ effort = ``true'' and its inverse were treated as a single combined rule. This aggregation captures refactoring co-occurrence across branches, regardless of ordering, thereby requiring recomputation of \textit{Lift}.

We analyzed 520 refactoring pairs that satisfied the minimum support threshold of 20 tuples. Most combinations exhibited \textit{Lift} values above 3, with an average of 5.17 (range 3.05--9.48), indicating that concurrent refactorings across branches are often associated with a higher frequency of merge effort. Notably, 54.0\% of the pairs presented \textit{Lift} values greater than 5.0, suggesting that this effect is widespread rather than limited to a small subset of combinations.

To obtain a comprehensive understanding of how refactoring combinations are associated with merge effort, we adopted a multidimensional analysis that integrates association strength (\textit{Lift}), effect size (Odds Ratio), statistical significance (Fisher’s Exact Test), and support of the rule. All reported p-values are below 0.0001. Given the large size of our dataset (91,270 tuples) and the imbalanced distribution of the effort variable (approximately 93\% of cases have effort = ``false''), statistical significance alone is insufficient to assess practical relevance. Therefore, we focused on the effect size (Odds Ratio) and the support of the rule to assess practical relevance. Based on these dimensions, we classify refactoring pairs into two categories according to their effect magnitude and robustness.

\textbf{Rare but high-impact combinations} correspond to pairs with very high effect magnitude but low support (Odds Ratio $\geq$ 10, high \textit{Lift}, and support $\geq$ 20 and $\leq$ 50). In this group, the Odds Ratio~(OR) threshold captures extreme effect sizes, indicating that the frequency of merge effort is at least ten times higher when these refactoring pairs occur compared to when they do not. At the same time, the support range indicates that these combinations are relatively infrequent in the dataset, meaning that their impact is concentrated in a limited number of merge scenarios. Although not among the most frequent patterns, these combinations exhibit stable and extreme associations with merge effort. In total, we identified 10 refactoring pairs that satisfy these criteria. 

Representative examples include \textit{Inline Variable} with \textit{Split Variable} (\textit{Lift} = 7.14, OR = 12.69, support = 49), \textit{Merge Attribute} with \textit{Split Attribute} (\textit{Lift} = 6.88, OR = 11.84, support = 45), \textit{Push Down Attribute} with \textit{Split Attribute} (\textit{Lift} = 6.82, OR = 11.65, support = 37), and \textit{Extract Subclass} with \textit{Split Attribute} (\textit{Lift} = 6.64, OR = 11.52, support = 29). These combinations often involve structural transformations that fragment, merge, or reorganize shared data structures, creating semantically incompatible changes across branches. In particular, many of these patterns involve operations that alter data representation or redistribute responsibilities across program elements, which can increase the chances of inconsistencies when applied concurrently in parallel branches.

\textbf{Very frequent and high-impact combinations} correspond to pairs that jointly exhibit high effect magnitude and very high support (Odds Ratio $\geq$ 7 and support $\geq 500$), typically accompanied by consistently high \textit{Lift} values. In this group, the Odds Ratio threshold identifies combinations with strong, stable effects, while the support constraint captures highly recurrent patterns in the dataset. Unlike rare high-impact combinations, these patterns occur frequently in our dataset, making them particularly relevant in practice.

Across this group, we identified 28 refactoring pairs involving 16 distinct refactoring types out of the 33 considered in this study. These types can be grouped into three main categories: (i) structural transformations (e.g., \textit{Extract Class}, \textit{Move Method} and \textit{Pull Up Attribute}), which modify program organization and class hierarchies; (ii) interface and contract changes (e.g., \textit{Rename Method}, \textit{Change Parameter Type} and \textit{Parameterize Variable}), which affect method signatures and data-structure representations; and (iii) local implementation adjustments (e.g., \textit{Inline Method} and \textit{Replace Variable With Attribute}), which operate at a finer granularity. Representative examples include \textit{Extract Class} with \textit{Extract Class} (\textit{Lift} = 5.67, OR = 8.79, support = 521), \textit{Move Attribute} with \textit{Move Class} (\textit{Lift} = 5.09, OR = 7.70, support = 508), \textit{Extract Superclass} with \textit{Rename Attribute} (\textit{Lift} = 5.01, OR = 7.55, support = 538), and \textit{Move Class} with \textit{Rename Class} (\textit{Lift} = 4.94, OR = 7.48, support = 606).

Notably, the most frequent and high-impact combinations are predominantly formed by interactions between structural transformations and interface-level changes. This suggests that merge effort is associated with concurrent modifications that alter program structure and exposed interfaces. Such combinations tend to produce semantically related but structurally incompatible changes across branches, which may make automatic resolution more challenging. This result suggests that a relatively small subset of refactoring types accounts for a large share of the most impactful and recurrent merge scenarios.

Overall, these results indicate that merge effort is associated with the interaction of concurrent structural changes across branches, rather than isolated refactoring operations. This pattern holds for both rare but high-impact and very frequent combinations, suggesting that similar interaction mechanisms occur across both extreme and recurrent scenarios. The most critical cases arise when transformations modify shared data representations, method signatures, or class hierarchies, producing semantically related but structurally incompatible changes that are difficult to reconcile automatically. However, the presence of a refactoring pair across parallel branches does not necessarily imply that both refactoring types contributed to the same regions of code in conflict. Refactoring pairs are not isolated within merge scenarios, as multiple other refactoring types and code changes often coexist in the same branches.

To illustrate how the interaction of refactorings across parallel branches can lead to merge effort, we present a concrete example of a textual conflict involving two distinct refactoring types. Code~\ref{cod:MergeEffortExample} shows a merge scenario identified in commit \texttt{e82e72d} from the \textit{OpenRefine} project. In this example, two branches introduced different refactorings to the same method. One branch applied a \textit{Rename Parameter}, changing the parameter name from \textit{totalCount} to \textit{rowCount}, while the other applied a \textit{Change Parameter Type}, modifying its type from \textit{long} to \textit{int}. During the merge, the developer manually reconciled these conflicting transformations by preserving the renamed parameter introduced in one branch while keeping the original \texttt{long} type, thus rejecting the change to \texttt{int}. This example highlights how even simple, common refactorings, when applied concurrently, can lead to merge effort that is not fully resolved automatically by merge tools.

\begin{listing}[htbp]
\caption{Example of merge effort identified in commit \texttt{e82e72d} of the \textit{OpenRefine} repository.}
\label{cod:MergeEffortExample}

\begin{lstlisting}[
    basicstyle=\ttfamily\scriptsize,
    escapeinside=@@,
    columns=fullflexible,
    keepspaces=true,
    aboveskip=0pt,
    belowskip=0pt
]
@\colorbox{diffdel}{\makebox[\linewidth][l]{\ttfamily\scriptsize - \ public void setRowCount(long totalCount) \{}}@
@\colorbox{diffadd}{\makebox[\linewidth][l]{\ttfamily\scriptsize + \ public void setRowCount(long rowCount) \{}}@
@\colorbox{diffdel}{\makebox[\linewidth][l]{\ttfamily\scriptsize \ -    public void setRowCount(long totalCount) \{}}@
@\colorbox{diffadd}{\makebox[\linewidth][l]{\ttfamily\scriptsize \ +    public void setRowCount(int totalCount) \{}}@
@\colorbox{diffadd}{\makebox[\linewidth][l]{\ttfamily\scriptsize ++ public void setRowCount(long rowCount) \{}}@
\end{lstlisting}

\end{listing}

\MyBox{\textbf{Answer to RQ3}: Merge effort is associated with the co-occurrence of refactoring types across parallel branches, rather than with isolated transformations. We observed two complementary patterns: rare but high-impact combinations, typically involving structural transformations that fragment or redistribute program elements (e.g., \textit{Split Variable}, \textit{Split Attribute}), often combined with operations that reorganize or consolidate structure (e.g., \textit{Merge Attribute}); and very frequent and high-impact combinations, characterized by interactions between structural reorganization (e.g., \textit{Extract Class}, \textit{Move Method}) and changes to method signatures and data-structure representations (e.g., \textit{Rename Method}, \textit{Change Parameter Type}, \textit{Parameterize Variable}). Overall, these results indicate that merge effort is primarily associated with concurrent refactorings, although this relationship should be interpreted as associative rather than causal.}

\subsection{\textbf{Discussion and Implications}}
\label{sec:resultsDiscussion}

This section interprets the results and presents practical implications.

Our results can be explained through concrete code-level merge scenarios. When refactorings occur in parallel and affect the same code region, structural transformations such as \textit{Move Class}, \textit{Move Method}, and \textit{Move Attribute} often induce textual conflicts if one branch relocates code while another modifies its original location, leading to overlapping deletions and edits. Similarly, \textit{Rename Attribute} and \textit{Change Return Type} may conflict with parallel edits that still reference previous identifiers or types, producing inconsistent method signatures or field usages. 

Even localized transformations, such as \textit{Extract Variable}, \textit{Inline Variable}, and \textit{Split Parameter}, can introduce conflicts when one branch rewrites expressions while the other modifies them. Such conflicts do not require refactorings in both branches, but often arise from interactions between refactorings and regular edits. When refactorings are applied in both branches, combinations such as \textit{Move Attribute + Move Class} may produce incompatible structural reorganizations, while combinations like \textit{Extract Superclass + Rename Attribute} and \textit{Move Class + Rename Class} introduce inconsistencies in class hierarchies and naming, generating conflicts in shared structures (e.g., class definitions and method signatures) that cannot be automatically merged.

However, not all merge effort is explained by textual conflicts. In some cases, refactorings applied in parallel branches can lead to semantically inconsistent code that compiles or merges without explicit conflicts but requires manual adjustments during integration (e.g., adapting method calls, updating dependencies, or reconciling partially moved logic). These situations result in additional lines added or removed in the merge commit, reflecting the effort required to restore functional correctness rather than resolve direct textual conflicts.

To contextualize our refactoring-based findings, we examined two baseline indicators of code changes: \textit{number\_lines\_changed} and \textit{number\_files\_changed} across the merging branches. We analyzed these attributes using the same association rule mining procedure described in Section~\ref{sec:association_rules}, based on the Apriori algorithm and the \textit{Lift} measure, with the same minimum support threshold of 20 merge commits.

To enable rule extraction, both metrics were discretized into quartile-based ranges: low ($>$0 and $\leq Q1$), medium ($>Q1$ and $\leq$ median), high ($>$ median and $\leq Q3$), and very high ($>Q3$). In our dataset, these correspond to: for \textit{number\_lines\_changed}, low ($>$0 and $\leq$ 175), medium ($>$175 and $\leq$ 975), high ($>$975 and $\leq$ 5,817), and very high ($>$5,817); and for \textit{number\_files\_changed}, low ($>$0 and $\leq$ 9), medium ($>$9 and $\leq$ 30), high ($>$30 and $\leq$ 125), and very high ($>$125).

When merge effort is analyzed as a binary attribute (\textit{effort = ``true''}), both metrics exhibit a consistent pattern: \textit{low} values are associated with substantially lower frequencies of merge effort, while \textit{very high} values are associated with substantially higher frequencies. For example, \textit{number\_lines\_changed} yields \textit{Lift} values ranging from 0.16 (-84\%) in the \textit{low} range to 2.50 (+150\%) in the \textit{very high} range, while \textit{number\_files\_changed} ranges from 0.18 (-82\%) to 2.45 (+145\%).

We then analyzed effort intensity using the same discretization adopted in RQ1 (\textit{low}, \textit{medium}, and \textit{high}). Under this setting, \textit{very high} values of \textit{number\_lines\_changed} yield \textit{Lift} values of 2.22 (+122\%), 2.64 (+164\%), and 2.98 (+198\%) for \textit{low}, \textit{medium}, and \textit{high} effort, respectively, while \textit{number\_files\_changed} yields corresponding values of 2.21 (+121\%), 2.56 (+156\%), and 2.85 (+185\%). These results show that larger change volumes are associated with both the occurrence of merge effort and higher effort intensity.

These findings confirm basic change volume metrics capture relevant aspects of merge complexity. In particular, higher numbers of changed lines and files are consistently associated with higher merge effort frequencies and intensities. However, when contrasted with our refactoring analysis, an important distinction emerges. While churn-related metrics reach \textit{Lift} values around 2.5--3.0, several refactoring types and combinations exhibit substantially stronger associations. For example, major \textit{Rename Attribute} refactorings reach \textit{Lift} values above 5.0, and certain concurrent refactoring combinations exceed \textit{Lift} values of 8.0. These results imply that refactoring-aware signals capture aspects of merge complexity that goes beyond the sheer volume of changes, providing a fine-grained perspective on how code transformations relate to merge effort~---~a motivational insight for the practical guidelines discussed next.

\textbf{Practical Guidelines.} Our findings suggest that refactorings, particularly structural ones and their concurrent application across branches, are often associated with increased merge effort. These results highlight the importance of coordination strategies, such as reducing branch divergence, integrating changes more frequently, and improving communication around refactoring activities. From a tooling perspective, they also indicate opportunities for refactoring-aware merge strategies and recommendation systems that consider refactoring type, volume, diversity, and cross-branch combinations.

To operationalize these implications, Table~\ref{tab:practical_guidelines} summarizes a set of actionable guidelines derived from RQ1--RQ3. The table organizes recurring refactoring scenarios, links them to recommended actions, and illustrates how to implement these recommendations in practice through pull request analysis, continuous integration pipelines, and repository-level tooling (e.g., GitHub Actions, Gerrit hooks, and bots) to support more informed merge decisions and reduce integration complexity. While each guideline is associated with a specific observed scenario, many recommendations are inherently cross-cutting and applicable to multiple contexts. We organize them according to their most representative scenarios to maintain clarity and alignment with our empirical findings.

\newcommand{\PracticalGuidelinesTable}{
\begingroup
\scriptsize
\setlength{\tabcolsep}{3pt}
\renewcommand{\arraystretch}{0.88}
\begin{tabular}{p{2.2cm} p{6.0cm} p{9.0cm}}
\toprule
\textbf{Observed Scenario} & \textbf{Practical Guideline} & \textbf{Tooling/Workflow Implementation (e.g., GitHub, Gerrit, CI)} \\
\midrule
\vspace{-2mm}

High volume of refactorings (High TNR) &
Consider splitting the refactoring into smaller increments or integrating changes earlier to reduce accumulation. &
\textbf{CI/CD Gate with Automated Bot Warnings}: Implement a GitHub Action with a rule-based bot that detects Pull Requests exceeding a predefined TNR threshold, automatically issuing warnings and suggesting earlier merges. The bot can also enforce mandatory senior reviews, acting as a CI/CD gate to mitigate risks associated with high volumes of refactorings. \\
\addlinespace

High diversity of refactoring types (High NDRT) &
Monitor integration risk as heterogeneous transformations increase overlapping edits. &
\textbf{Integration Assistant:} Deploy a bot to provide PR-level analytics on refactoring diversity, displaying risk indicators in GitHub checks or Gerrit dashboards\footnotemark[1] to alert reviewers about complex structural changes and support more informed review decisions.  \\
\addlinespace

Structural refactorings (e.g., Move Class, Push Down Method) &
Avoid concurrent applications across branches. Prioritize developer communication if unavoidable, and consider isolating large structural refactorings into dedicated branches when possible. &
\textbf{Structural Refactoring Conflict Predictor:} Use a Gerrit hook or GitHub bot integrated into CI pipelines to detect structural refactorings (e.g., via RefactoringMiner) and identify concurrent changes across Pull Requests that modify the same class hierarchy. The system can proactively alert developers about potential conflicts arising from overlapping structural changes at the repository level. \\
\addlinespace

Concurrent refactorings on related code elements &
Review Pull Requests for interactions between refactorings and parallel edits. Sequence changes to minimize interference, and promote early communication among developers when such changes are unavoidable. &
\textbf{Dependency Mapping with Cross-Branch Analysis:} Integrate refactoring detection techniques into CI pipelines to visualize impacted code regions directly in the Pull Request dashboard, while performing cross-branch analysis to detect overlapping edits or refactorings across active branches and proactively notify developers before integration (e.g., using tools such as RefactoringMiner). \\
\addlinespace

Presence of high-risk combinations (RQ3 Results) &
Increase attention during code review and merge preparation for known high-risk pairs (e.g., Split vs. Extract refactorings), with tool support that automatically flags these combinations and assists reviewers. &
\textbf{High-Risk Refactoring Linter/Bot with Escalation:} Deploy a rule-based bot (e.g., GitHub Actions) that detects high-Lift refactoring combinations, automatically labels Pull Requests as ``High Integration Risk'', and suggests safer structural alternatives. Additionally, the bot can trigger automated alerts in team communication channels (e.g., Slack\footnotemark[2] and Discord\footnotemark[3]) to support immediate coordination and risk-aware integration decisions. \\
\addlinespace

Refactoring-heavy branches vs. feature work &
i) Establish dedicated branches to isolate massive structural modifications. ii) Prioritize the integration of refactoring-heavy branches, prefer short-lived branches, and reduce prolonged parallel development on affected modules. iii)Consider adopting trunk-based development with feature toggles~\cite{trunkbased2022, rezvan2021} to reduce prolonged parallel divergence by enabling early integration and limiting the accumulation of changes.
&
\textbf{Refactoring-Aware Merge Control and Coordination:} Use Gerrit’s voting or GitHub’s ``Draft Pull Request'' status to temporarily block feature merges until structural refactoring is integrated, while leveraging dashboards that indicate branch divergence and refactoring density to support merge prioritization and coordination decisions. \\
\addlinespace

Refactoring-aware merge strategies and timing &
Consider merge strategies that separate edits from refactorings and schedule merges based on refactoring thresholds to reduce integration complexity. &
\textbf{Operation-Based Merge and Timing Assistant:} Implement merge workflows that first integrate non-refactoring edits and then replay refactorings (inspired by operation-based merging~\cite{mens2002}). Additionally, use CI-integrated tools to monitor refactoring counts across branches and recommend optimal merge timing, triggering alerts when thresholds indicate increased merge effort risk.\\
\addlinespace

\bottomrule
\end{tabular}
\endgroup
}

\begin{table*}[t]
\centering
\caption{Actionable Guidelines and Tooling Support Derived from Refactoring-Merge Effort Associations}
\label{tab:practical_guidelines}
\PracticalGuidelinesTable
\end{table*}

\footnotetext[1]{\url{https://www.gerritcodereview.com/}}
\footnotetext[2]{\url{https://slack.com/}}
\footnotetext[3]{\url{https://discord.com/}}

\vspace{-2mm}
\section{Threats to Validity}
\label{sec:threats}

This section discusses internal, external, construct, and conclusion threats that may have influenced the reported results.

\textbf{Internal Validity}. One potential threat concerns the completeness of the collected data on refactorings. We imposed a 10-minute \textit{RefactoringMiner} timeout, following literature recommendations~\cite{mahmoudi2019} to prevent analysis from hanging on long-running commits. In rare cases (only 94 commits, representing 0.0001\% of our dataset), the process terminated after exceeding this threshold. Although this is a negligible portion of the data, non-detected refactorings may pose a minor threat to the validity of aggregated results.

We analyzed co-occurrence patterns using association rules, interpreting them as dataset-specific associations rather than evidence of causality. This technique identifies frequent combinations based on support, confidence, and lift, enabling the detection of refactoring patterns associated with higher code churn during merges. However, it does not control for latent factors such as project size, branching policies, or system modularity, and relies on code churn as a proxy for merge effort, limiting causal inference and internal validity. To mitigate spurious conclusions, we conducted sensitivity analyses and stratified results by the total number of refactorings (TNR) and the number of different refactoring types (NDRT). Nonetheless, future studies could further investigate causal mechanisms and assess the generalizability of these findings in other contexts.

Discretizing numeric attributes into binary or categorical ranges, required by the Apriori algorithm, introduces an internal validity threat by reducing data granularity and creating arbitrary thresholds. Such discretization may homogenize distinct behaviors (e.g., 10 vs. 90 refactorings) and obscure subtle patterns. We mitigated this by using two strategies: a four-tier discretization (0, 1--9, 10--99, $\geq 100$) that aligns with natural data clusters and highlights rare, high-impact cases (RQ1), and a binary form (0 vs. $>$0) to simplify presence analysis and control rule volume (RQ2 and RQ3). While some precision is lost, the trade-off enabled interpretable pattern discovery at scale, and the study still revealed a substantial number of consistent associations that support the conclusions discussed in Section~\ref{sec:resultsDiscussion}. Future work can explore dynamic discretization techniques~\cite{lud2000,li2012,tan2018} to capture finer-grained patterns.

Although our merge effort metric does not focus exclusively on textual conflicts, developers may only detect build, test, or production conflicts after completing the merge. Ghiotto et al.~\cite{ghiotto2020} refer to these situations as postponed merges, in which developers resolve conflicts through changes committed after the merge commit. Nevertheless, their study found that such cases are rare, representing about 0.97\% of the analyzed merge scenarios, suggesting that postponed resolutions have a limited impact on our overall results.

\textbf{External Validity}. Our study focuses on open-source Java projects, and we do not claim that our findings generalize to all Java systems, particularly to closed-source enterprise software or to projects written in other programming languages. Selecting mature, well-established repositories may limit applicability to less mature projects that may exhibit different development dynamics. Additionally, restricting the dataset to primarily English-language repositories and excluding projects with extremely high or low commit counts may have reduced diversity, potentially underrepresenting certain refactoring and merge scenarios. Therefore, the generalizability of our findings is limited to the characteristics of the selected sample. Despite these constraints, our analysis revealed consistent behavioral patterns, suggesting that similar trends may exist in other contexts and warrant further investigation.

\textbf{Construct Validity}. To operationalize merge effort, we relied on code churn, measured as the number of lines added or removed during merge commits. This metric has been widely adopted in software engineering as a proxy for effort and change complexity~\cite{sjoberg2012, olsson2017, carka2022, jesse2023, bessghaier2025}. However, code churn captures only one observable dimension of merge effort, primarily reflecting the magnitude of code modifications required during integration, and may not fully represent other important aspects, such as cognitive load, coordination among developers, and the time required to resolve conflicts. As a result, scenarios involving large but straightforward changes (e.g., rename-based refactorings supported by automated tools) may be overestimated in terms of effort, while small but complex conflicts requiring careful reasoning may be underestimated. To mitigate this limitation, we explicitly frame code churn as a proxy for merge effort and interpret our results within this scope, while complementing our analysis with different perspectives on change (e.g., refactoring types, diversity, and combinations), which provide a more fine-grained view beyond change volume alone.

In this study, refactorings are identified in the branches' development history prior to merge commit. Thus, our analysis focuses on refactoring activities introduced before the integration step and does not include refactorings along merge resolution. Although developers may occasionally introduce refactorings while resolving conflicts, this scenario might be generally uncommon, as merge resolution typically focuses on integrating concurrent code changes rather than introducing new structural modifications. Consequently, excluding refactorings introduced during the merge operation may lead to a slight underestimation of refactoring activity in specific cases. This fact does not materially affect our results, as the study investigates associations between refactoring activities made in parallel branches and merge effort.

\textbf{Conclusion Validity}. We observe that some of the identified association rules show relatively low support, occurring in only 20 merge commits. As a result, some rules may have occurred by chance. However, removing them would eliminate rare yet potentially relevant patterns.

\vspace{-2mm}
\section{Related Work}
\label{sec:related}

Several studies explored the relationship between refactorings and merge operations~\cite{dig2007, laszlo2007, lebenich2017, mahmoudi2018, mahmoudi2019, ellis2023, oliveira2023}. Most~\cite{laszlo2007, dig2007, lebenich2017, ellis2023} focused on developing, improving, or proposing merge tools that account for refactorings in different branches. Only some studies~\cite{mahmoudi2019, oliveira2023, oliveira2023_b} adopted an analytical perspective, investigating the relationship between refactorings and merge conflicts. This section discusses existing contributions and research gaps.

Angyal et al.~\cite{laszlo2007} extended the three-way merge algorithm to support renames, incorporating them into change reconciliation. Dig et al.~\cite{dig2007} introduced \textit{MolhadoRef}, a tool identifying seven refactoring types before code merge using \textit{RefactoringCrawler}. This tool treats refactorings and edits as recorded/replayed change operations. They proposed inverting refactorings, performing a merge, then replaying refactorings to address intermingled edits and refactorings, comparing results against CVS~\cite{morse1996}. Leßenich et al.~\cite{lebenich2017} enhanced \textit{JDIME} to resolve conflicts involving specific refactorings like code movements. They proposed a syntax-aware structured merge strategy based on heuristic optimization and syntax-specific lookahead during tree matching. 

Ellis et al.~\cite{ellis2023} further investigated refactoring-aware merging by empirically evaluating operation-based approaches such as \textit{RefMerge} and \textit{IntelliMerge}. Their study analyzed 2,001 merge scenarios extracted from 20 Java open-source projects and evaluated how refactoring-aware algorithms perform compared to traditional merge strategies. Their findings show that refactoring-aware approaches can reduce or automatically resolve a portion of merge conflicts that conventional tools fail to handle. Unlike our work, their study focuses on evaluating merge algorithms rather than analyzing how different refactoring types relate to merge effort. In contrast, our study investigates more than 90,000 merge commits across 64 Java projects and focuses on identifying patterns between refactoring types, their cross-branch combinations, and merge effort using association rule mining.

Mahmoudi and Nadi~\cite{mahmoudi2018} investigated changes made by phone vendors compared with original Android development, focusing on concurrent evolution and change overlap. They reported common refactorings and examined the feasibility of automated merge support, considering a small subset of six frequent types. Extending this work, Mahmoudi et al.~\cite{mahmoudi2019} analyzed the relationship between 15 refactoring types and merge conflicts across 3,000 Java open-source projects. They defined ``involved refactoring'' as occurring in changes that overlap conflicting regions, and classified a merge scenario as involving refactorings when at least one conflicting region contains such a transformation. Their results showed that 22\% of merge conflicts involve refactorings, with 11\% of conflicting regions including at least one, and that \textit{Extract Method} appears more frequently in conflicts. Additionally, conflicting regions with refactorings tend to be larger and more complex, and such scenarios often involve more evolutionary changes.

While Mahmoudi et al.~\cite{mahmoudi2019} provide an analysis of the presence of refactorings in merge conflicts across 3,000 Java projects, their study focuses on identifying whether refactorings are involved in conflicting regions and on characterizing conflict properties, such as their size. In contrast, our study analyzes a curated dataset of 64 projects and investigates a different dimension of the problem by examining how specific refactoring types and their combinations relate to the effort required to resolve merges. From a methodological perspective, Mahmoudi et al. define ``involved refactoring'' based on overlap with conflicting regions, whereas we adopt an association-rule mining approach that combines \textit{Lift}, Odds Ratios, Fisher’s Exact Test, and support to analyze the relationship between refactoring types and merge effort. While their study considers 15 refactoring types, we extend this analysis with a more fine-grained investigation of 33 types. Mahmoudi et al. report that refactorings are present in a substantial portion of merge conflicts and that such conflicts tend to be larger and more complex. Our results go beyond presence and show that specific refactoring types, their volume, diversity, and cross-branch combinations, are associated with increased merge effort. In particular, we identify that structural refactorings and concurrent combinations across branches are more strongly associated with higher merge effort, highlighting patterns that are not captured solely based on conflict occurrence.

Building upon this methodology, Mongiovi et al.~\cite{oliveira2023_b}  extended the investigation to 50 JavaScript repositories (56,966 merge scenarios, 4,816 with conflicts) using RefDiff 2.0~\cite{silva2020refdiff}. They found that 7\% of merge scenarios have at least one refactoring within conflicting files, while 4\% involved refactoring actions at the conflict region level. They observed a positive but weak correlation between these variables at both levels, and a moderate positive correlation between refactoring type diversity and merge scenarios with conflicts. Among the various refactoring types, ``Internal Move'' and ``Move'', which encompass types such as Move Attribute, Move Variable, Move Method, Move Class, and Extract Method, were most commonly associated with merge conflicts. 

Previous studies investigated the phenomena of composite refactorings~\cite{bibiano2024, bibiano2021}. Composite refactorings are interlinked transformations that consists of two or refactorings (e.g., \textit{two Extract Methods} or \textit{one Extract Method + one Move Method}. A composite refactoring is considered complete when it fully removes the intended problem code smell)~\cite{BibianoICPC2020}. Incomplete refactorings occur as often as complete refactorings~\cite{bibiano2021}. The incompleteness may occur as developers aimed at reducing merge effort later. Existing studies show that developers roll back refactorings~\cite{PaixaoMSR2020, UchoaICSME20}, which may be induced by reduction of merge effort. However, those authors did not study the relationship between refactorings and merge effort. 

Taken together, these prior studies highlight the importance of understanding how refactorings influence merge operations and the need to improve refactoring-merge-aware tools. However, most of these studies focus on a small subset of refactoring types, despite the wide variety of transformations commonly observed in real-world software development~\cite{fowler2018, bibiano2021, alameer2026, liu2025, wang2025}. Additionally, they tend to examine the relationship between specific refactorings and the occurrence of merge conflicts, with an emphasis on conflict chunk size rather than the effort required to resolve these conflicts. 

By addressing these combined limitations, our work deepens the analysis by investigating not only the specific behavior of each refactoring type but also how different combinations, applied simultaneously in parallel branches, relate to merge effort. In doing so, it complements and extends previous findings, offering a more fine-grained understanding of the mechanisms through which individual refactorings and their interactions increase the complexity and cost of integrating parallel development changes.

\vspace{-2mm}
\section{Conclusion}
\label{sec:conclusions}

Refactorings are central to improving code quality, but can introduce merge risks that are often overlooked in collaborative development. Our results show that the presence, frequency, and diversity of certain refactoring types, particularly renamings and structural changes (e.g., move, extract, split), are associated with increased merge effort, whereas more localized modifications (e.g., parameter or variable adjustments, attribute or variable extractions) exhibit weaker associations. We also find that combinations of refactorings across branches tend to show stronger associations with merge effort, especially when involving structural transformations that reorganize elements. These findings require better means for coordination when applying high-impact refactorings, as well as tool support to detect and mitigate merge risks.

\textbf{Key Lessons Learned.} Our findings indicate that: (1) change volume (e.g., number of modified lines and files) provides a useful baseline but is insufficient alone, as refactoring-aware signals capture structural aspects of change not explained by churn; (2) refactoring type plays a central role, with structural transformations showing consistently stronger associations with merge effort, even at low volumes; (3) refactoring volume (TNR) and diversity (NDRT) are independent drivers of merge complexity; (4) the concurrent application of refactorings across branches is a critical risk factor, particularly when combinations modify shared data-structure representations, method signatures, or redistribute responsibilities across code elements; (5) localized transformations tend to have more limited impact on integration than architectural changes affecting class hierarchies and shared abstractions; (6) refactoring interactions amplify merge effort, even for transformations that are typically safe in isolation; (7) merge effort emerges not only from textual conflicts but from the interaction of concurrent changes, reinforcing the need for coordination strategies and refactoring-aware tool support; and (8) merge effort should be studied from a multidimensional perspective, as factors such as cognitive load and resolution time motivate further qualitative investigation for deeper understanding.

Our findings lay the groundwork for building refactoring risk models and promoting safer collaboration in parallel development workflows. Future work may include validating these insights in industrial settings, exploring other programming languages, including additional refactoring types not explored in this study, conducting qualitative studies with developers, and improving semantic conflict detection to deepen understanding of how refactorings relate to merge effort. Additionally, future work may explore alternative effort proxies, such as fine-grained change measures based on Levenshtein edit distance~\cite{yujian2007}, to complement line-based churn metrics and provide a more detailed view of code modifications. We also consider relying on eyer trackers to derive measurements of cognitive effort. The analysis of eye movements while a developer is performing the merge would capture elements associated with the cognitive process. Recent studies has used eye trackers to assess the cognitive effort of developers understanding refactored code~\cite{aldoJSS2026, aldoEMSE2021} and opportunities for refactoring (smells)~\cite{martinsSBES2024}. No existing work has explored the use eye trackers on monitoring merge effort.

Finally, we consider it an important direction to investigate how AI-assisted development tools may influence merge effort, particularly in scenarios where automated support can reduce manual intervention, and how AI agents may affect the relationship between code changes and perceived effort in modern collaborative development practices. Another potential direction for future work is investigating refactorings introduced during merge commits. It would be interesting to analyze whether they primarily serve to reconcile conflicting changes or whether developers use the merge process to opportunistically improve the code.

\vspace{-2mm}
\section*{Acknowledgments}

We acknowledge the use of ChatGPT~5 and Grammarly for improving the grammar, vocabulary and text style. All suggestions were carefully examined and fixed by the authors; we take full responsibility for the paper  content. The authors would like to thank CNPq (grant 315711/2020-5, 309300/2023-1, 420025/2023-5, 310391/2025-3, and 308984/2025-0) and FAPERJ (grant E-26/201.139/2022, E-26/200.510/2023, E-26/211.033/2019, E-26/211.033/2019 and E-26/204.145/2024) for the financial support.

\vspace{-2mm}
\bibliographystyle{IEEEtran}
\bibliography{TSE2025_MergeRefactoring}

\vfill

\end{document}